\documentclass[twoside,leqno,twocolumn]{article}

\usepackage{caption}
\usepackage{subcaption} 

\usepackage{lmodern}\usepackage[T1]{fontenc} 
\usepackage[utf8x]{inputenc}
\usepackage[dvipsnames, svgnames, x11names]{xcolor}
\usepackage{todonotes}
\usepackage{authblk}

\usepackage{listings}
\usepackage{tikz}
\usetikzlibrary{positioning}
\usepackage{pgfplots}
\pgfplotsset{compat=1.13}

\usepackage{booktabs}

\usepackage[margin=0.92in]{geometry}

\usepackage{rotating}

\usepackage[numbers, sort&compress]{natbib}
\usepackage[english]{babel}

\usepackage{amsmath, amssymb, mathtools}
\allowdisplaybreaks
\usepackage{oubraces}

\usepackage{nicefrac}
\usepackage{grffile} 
\usepackage{enumitem} 
\usepackage{tabularx} 
\usepackage{longtable}

\usepackage{pdfpages}

\newcommand{\repEq}{\overset{r}{=}}
\newcommand{\repLess}{\overset{r}{<}}
\newcommand{\iso}{\ensuremath{\cong}}

\newcommand{\moreSpecial}{\ensuremath{\preceq}}

\usepackage{siunitx}
\usepackage{tikz}
\usepackage{chemfig,chemmacros}
\usetikzlibrary{arrows, calc, decorations, decorations.pathmorphing, matrix, shapes}

\makeatletter 
\tikzset{position/.style args={#1 degrees from #2}{
		at=(#2.#1), anchor=#1+180, shift=(#1:\tikz@node@distance)
}}
\tikzset{%
	name prefix/.code={%
		\tikzset{
			name/.code={\edef\tikz@fig@name{#1 ##1}}
		}%
	}%
}
\makeatother

\tikzset{every matrix/.style={ampersand replacement=\&}}

\tikzset{eMorphism/.style={->, >=stealth}}
\tikzset{eIsomorphism/.style={double, double equal sign distance}}

\tikzset{vDot/.style={fill,minimum size=0, inner sep=1pt, circle, outer sep=0}}
\tikzset{eDot/.style={}}
\tikzset{eDotMorphism/.style={->, >=stealth, dashed, red}}
\tikzset{gDot/.style={node distance=1em and 1em}}
\tikzset{gCat/.style={node distance=2em and 2em}}
\tikzset{vCat/.style={draw, minimum size=2.5em}}

\usepackage[noend,ruled,noline,linesnumbered,algo2e]{algorithm2e}
\DontPrintSemicolon
\SetKw{Break}{break}
\SetKw{Abort}{abort}
\SetKw{Continue}{continue}
\SetKw{Yield}{yield}
\SetKw{Nil}{nil}
\SetKw{And}{and}
\SetKw{Or}{or}
\SetKw{Not}{not}
\SetKw{False}{false}
\SetKw{True}{true}
\makeatletter
\newcommand{\removelatexerror}{\let\@latex@error\@gobble}
\makeatother

\newcommand\partition[1]{\pi_{#1}}

\DeclareMathOperator{\Aut}{Aut}
\DeclareMathOperator{\Stab}{Stab}
 
\DeclareMathOperator*{\argmin}{arg\,min}

\newcommand\email[1]{\texttt{#1}}

\usepackage{hyperref}
\usepackage{marvosym}
\newcommand\corr{\text{ (\Letter)}}

\title{\Large A Generic Framework for Engineering\\ Graph Canonization Algorithms}
\author[1-4]{Jakob L.\ Andersen\corr}
\author[1]{Daniel Merkle}
\affil[1]{Department of Mathematics and Computer Science, University of Southern Denmark, Odense M DK-5230, Denmark
	\email{daniel@imada.sdu.dk}}
\affil[2]{Research group Bioinformatics and Computational Biology, Faculty of Computer Science, University of Vienna, 1090 Vienna, Austria
	\email{jakob.lykke.andersen@univie.ac.at}}
\affil[3]{Institute for Theoretical Chemistry, University of Vienna, 1090 Wien, Austria}
\affil[4]{Earth-Life Science Institute, Tokyo Institute of Technology, Tokyo 152-8550, Japan}
\date{}

\begin{document}
\maketitle

\begin{abstract}
The state-of-the-art tools for practical graph canonization are all based on the indivi\-dualization-refinement paradigm,
and their difference is primarily in the choice of heuristics they include and in the actual tool implementation.
It is thus not possible to make a direct comparison of how individual algorithmic ideas affect the performance on different graph classes.

We present an algorithmic software framework that facilitates implementation of heuristics as independent extensions to a common core algorithm.
It therefore becomes easy to perform a detailed comparison of the performance and behaviour of different algorithmic ideas.
Implementations are provided of a range of algorithms for tree traversal, target cell selection, and node invariant, including choices from the literature and new variations.
The framework readily supports extraction and visualization of detailed data from separate algorithm executions for subsequent analysis and development of new heuristics.

Using collections of different graph classes we investigate the effect of varying the selections of heuristics,
often revealing exactly which individual algorithmic choice is responsible for particularly good or bad performance.
On several benchmark collections, including a newly proposed class of difficult instances, we additionally find that our implementation performs better than the current state-of-the-art tools.
\end{abstract}

\section{Introduction}
Graph canonization is the process of finding a canonical representation of a graph, such that all isomorphic graphs are assigned the same representation.
The graph isomorphism problem can thus be reduced to comparing such canonical representations,
which especially is useful when we want to test isomorphism against a large collection of graphs, e.g., for database querying.
There is a rich literature on the complexity of both graph canonization and graph isomorphism.
For longer discussion we refer to \cite{pgi:2}, and simply note that for general graphs the problems are not known to be NP-complete,
and the best bound for canonization is currently $e^{O(\sqrt{n\log n})}$ \cite{Babai:1983,BabaiHandbook},
while a quasi-polynomial bound for isomorphism was recently presented \cite{babai}.

For practical graph canonization there has also been extensive work, with several competitive tools being published in the last decades.
They are all build on the same core idea of a tree search over gradually more refined partitions of the vertex set, also called the \emph{individualization-refinement paradigm}.
Their difference is thus essentially in the heuristics for traversing and pruning the search tree, and how partitions are being refined.
One of the most successful tools is nauty \cite{pgi:1}, which not only finds a canonical representation but also computes the automorphism group,
which during the canonization is used for pruning the search tree.
Later tools, Bliss \cite{bliss,bliss:2} and Traces \cite{pgi:2}, also use this technique with the latter introducing a new way to exploit the discovered automorphisms.
A related tool is Saucy \cite{saucy:1,saucy:2} which only performs computation of the automorphism group, for which it introduced new heuristics to discover them.
Similarly there is Conauto \cite{conauto} which performs isomorphism testing directly without computing a canonical form.
Each new tool and updated versions of tools has incorporated ideas from the other tools, with further development of heuristics.
However, the performance comparisons have been done between the tools in their entirety,
making it exceedingly difficult to discover exactly which combination of ideas lead to better or worse performance on particular classes of graphs.
As the tools are almost completely independent implementations it is additionally hard to make a fair comparison, even if individual innovations could be isolated.

To address these problems we have developed a generic framework for constructing variations of graph canonization algorithms,
where new ideas can be implemented as separate plugins, and injected into a common core algorithm.
Not only does this framework solve the problem of fairly comparing heuristics,
but it also significantly lowers the barrier of entry for people to test new ideas in practice.
We provide implementations of a core set of heuristics, including new variations of node invariants and a memory sensitive tree traversal algorithm.
Contrary to the established tools the framework allows for direct canonization of graphs with edge attributes,
and we have developed a generalization of the widely used Weisfeiler-Leman refinement function which can exploit such attributes.

Using established benchmark graphs we discover interesting performance differences among combinations of heuristics,
including combinations with significantly different scaling behavior and better performance than the established tools.
For a recently proposed collection of six difficult graphs classes \cite{cfi-rigid,web:cfi-rigid}
we perform a detailed benchmark of the effects of node invariants which suggests why some of them are more difficult than the others.
Plots for all benchmarks can be found in the GitHub repository \cite{gh:graphCanon}.

The framework is implemented as a C++ library, called \textbf{GraphCanon}, with heavy use of generic programming \cite{genericProgramming} with influences from and compatibility with the Boost Graph library \cite{BGL,GGCL}.
It is available on GitHub \cite{gh:graphCanon} along with the accompanying \textbf{PermGroup} library \cite{gh:permGroup} for handling permutation groups.
In App.~\ref{app:code} we provide additional details of the framework, including pseudocode with direct links to the corresponding C++ code.
The appendix also contains additional visualizations of search trees and data from experiments.

In Sec.~\ref{sec:predef} and \ref{sec:abstractAlg} we lay out the mathematical description of the individualization framework,
setting the stage for the description of the framework in Sec.~\ref{sec:framework}.
The experimental results are presented in Sec.~\ref{sec:results}, and a summary with future developments is in Sec.~\ref{sec:conclusion}.

\section{Preliminary Definitions}
\label{sec:predef}

We denote an undirected graph as $G = (V, E)$, with $V$ as the vertices and $E$ as the edges.
The goal is to find a canonical representation of $G$, and we therefore assume the vertices already to have associated IDs.
For ease of notation we assume $V = \{1, 2, \dots, n\}$.
An attributed graph is a tuple $G = (V, E, l_V, l_E)$ of a graph $(V, E)$ and two attribution functions $l_V\colon V\rightarrow \Omega_V$ and $l_E\colon E\rightarrow \Omega_E$.
We assume that the attribute sets $\Omega_V$ and $\Omega_E$ are totally ordered sets.
We denote the set of all attributed graphs on $n$ vertices as $\mathcal{G}_n$ or simply as $\mathcal{G}$.
For the remainder of this contribution we assume an attributed graph $G = (V, E, l_V, l_E)$ with $n$ vertices is given.
Note that in related works, e.g., \cite{bliss} and \cite{pgi:2}, they use so-called colored graphs that are equivalent to graphs only with vertex attributes,
and they do not consider edge attributes directly.
We assume the set of graphs $\mathcal{G}$ is totally ordered.
For example, if the graphs are represented as adjacency matrices we can lexicographically compare the matrices.%
\footnote{An illustrated example of comparison using adjacency lists can be found in App.~\ref{app:adjRep}.}
For graphs with vertex and/or edge attributes this comparison must also account for those attributes.
For two graphs $G_1, G_2\in \mathcal{G}$ with the same underlying representation we say they are \emph{representationally equal}, written $G_1 \repEq G_2$.
When they are not we may say that one is \emph{representationally smaller}, written $G_1\repLess G_2$.

A canonization algorithm starts with the assumption that all vertices are unordered, and then incrementally introduces order.
To represent these intermediary partial orders we use the following construct.
An \emph{ordered partition} of $V$ is a sequence $\pi = (W_1, W_2, \dots, W_r)$ of non-empty sets of vertices that partitions $V$.
Each of the constituent vertex sets is called a \emph{cell} of $\pi$.
For a vertex $v$ in the $j$-th cell, we define $cell(v, \pi) = j$.
A cell of size 1 is called a \emph{singleton}, and if all cells are singletons
we call the partition \emph{discrete}.
If the ordered partition only has one cell, i.e., $\pi = (V)$, it is called the \emph{unit partition}.

The set of all ordered partitions over $V$ is denoted $\Pi$.
An ordered partition $\pi'$ is \emph{at least as fine as} $\pi$, written $\pi'\moreSpecial\pi$,
when $cell(u, \pi) < cell(v, \pi)$ implies $cell(u, \pi') < cell(v, \pi')$ for all $u,v\in V$. 
That is, $\pi'$ can be obtained from $\pi$ by only subdividing cells.

Ordered partitions are used to represent intermediary states of the canonization procedure,
in the sense that for a partition $\pi$ the vertices of a cell $W_i$ are said to be ordered before vertices of a cell $W_j$ when $i < j$.
The unit partition thus represents no ordering information, while each discrete ordered partition is a canonical order candidate.

Let $S_n$ denote the symmetric group on the set of vertices $V$.
For a permutation $\gamma\in S_n$ we denote the image of an element $v\in V$ as $v^\gamma$.
Composition of permutations is thus written from left to right, i.e., $v^{\gamma_1\gamma_2} = (v^{\gamma_1})^{\gamma_2}$.
The inverse of a permutation $\gamma$ is denoted $\overline{\gamma}$.
The permutation of a subset of vertices $W\subseteq V$ with $\gamma\in S_n$ is defined as $W^\gamma = \{w^\gamma\mid w\in W\}$,
while the permutation of a sequence $Q = (q_1, q_2, \dots, q_k)\in V^k$ is $Q^\gamma = (q_1^\gamma, q_2^\gamma, \dots, q_k^\gamma)$.
Similarly we extend permutation to combinations of these structures, all derived from $V$,
which in particular means we can permute ordered partitions and (representations of) graphs.

We interpret a discrete ordered partition $\pi$ as a permutation in $S_n$, that maps each cell index to its contained vertex.
That is, if $cell(v, \pi) = j$ then $j^\pi = v$.
The inverse permutation $\overline{\pi}$ thus maps vertices to their cell indices.
Note that if we use a discrete ordered partition $\pi$ to represent a candidate for the canonical order,
we then have $\pi$ as a permutation from the candidate canonical indices to the indices in the input graph.
Permuting the input graph with the inverse permutation $G^{\overline{\pi}}$ thus gives us the actual candidate for the canonical representation.

Two graphs $G_1, G_2\in \mathcal{G}$ are isomorphic, denoted $G_1\iso G_2$ if
there exists a permutation $\gamma\in S_n$ such that $G_1^\gamma \repEq G_2$.
The permutation $\gamma$ is then called an \emph{isomorphism}, and if $G_1$ and $G_2$ refer
to the same object, it is further called an \emph{automorphism}. 
The set of all automorphisms of a graph $G$, the automorphism group, is denoted $\Aut(G)$.

A canonization algorithm can be seen as a function on graphs, $C\colon\mathcal{G}\rightarrow \mathcal{G}$,
with the following properties, \cite[C1 and C2]{pgi:2}:
$C(G)\iso G$ and $C(G^\gamma) \repEq C(G)$, for all $\gamma \in S_n$.
That is, it returns a graph isomorphic to its input, and it is invariant with respect to permutations of its input.
The second property is also called \emph{isomorphism invariance},
and we will require this property for most of the procedures we describe in the following sections.

\section{The Abstract Algorithm}
\label{sec:abstractAlg}
For a high-level description of the individualization-refinement
approach with proofs of correctness we refer to \cite{pgi:2}.  The
following description follows the same principles, but we opt for a
description that more easily maps to the generic implementation
presented in the next section.

The individualization-refinement approach is a tree search starting
from the unit partition in the root
with each leaf of the search tree corresponding to a discrete ordered partition.
The canonical form then corresponds to a ``minimum'' leaf,
where ``minimum'' is defined in conjunction by the graph representation comparator $\repLess$,
and by so-called \emph{node invariants} that will be discussed in the end of this section.
Instead of comparing the vertex attributes in the leaves using $\repLess$ we can instead exploit them from the beginning by starting with a specific ordered partition instead of the unit partition.
The \emph{initial partition} is then $\pi_0 = (V_1, V_2, \dots, V_k)$ with the property that for all $u\in V_i$ and $v\in V_j$,
if $l_V(u) = l_V(v)$ then $i = j$, otherwise $l_V(u) < l_V(v)$ implies $i < j$.
That is, the vertices are partitioned by their attribute, and the cells are ordered by the attribute.

The algorithm is defined using several abstract functions.
The first one is a \emph{refinement function} $R\colon \mathcal{G}\times \Pi\rightarrow \Pi$,
with the properties \cite{pgi:2}: $R(G, \pi)\moreSpecial \pi$ and $R(G^\gamma, \pi^\gamma) = R(G, \pi)$ for all $\gamma\in S_n$.
That is, it produces a partition that is finer or equal to its input, and it is isomorphism invariant.
When the refinement function produces non-discrete partitions we must decide on a cell where we will artificially introduce cell splits.
For this we need a \emph{target cell selector} $T\colon \mathcal{G}\times \Pi\rightarrow 2^V$
that for a non-discrete partition returns one of the non-singleton cells.
This function must also be isomorphism invariant, i.e., $T(G^\gamma, \pi^\gamma) = T(G, \pi)^\gamma$, for all $\gamma\in S_n$.
The introduction of artificial splits is done by \emph{vertex individualization}.
For an ordered partition $\pi$ with a non-singleton cell $W$ and a vertex $v\in W$,
we define $\pi\downarrow v$ as the ordered partition where $W$ is replaced by two cells $\{v\}$ and $W\backslash\{v\}$, in that order.
That is $\pi\downarrow v$ is the partition strictly finer than $\pi$ obtained by individualizing $v$ to the left from the rest of its cell.

\paragraph{Canonization as a Tree Search}
A search tree can now be formally defined as follows.
Each node $\tau$ of the tree is identified by a sequence of vertices,
and it implicitly defines an associated ordered partition $\partition{\tau}$ defined in the following manner.
Let $\pi_0$ be the initial ordered partition constructed from vertex attributes as described above.
The root of the search tree is then the empty sequence $\tau_{root} =()$,
with the associated ordered partition $\partition{\tau_{root}} = R(G, \pi_0)$.
For a node $\tau = (v_1, v_2, \dots, v_k)$,
if $\partition{\tau}$ is discrete then $\tau$ is a leaf.
Otherwise, let $W = T(G, \partition{\tau})$ be the target cell selected by $T$.
For each vertex $w\in W$ there is a child node $\tau_{child} =  (v_1, v_2, \dots, v_k, w)$,
with the associated ordered partition $\partition{\tau_{child}} = R(G, \partition{\tau}\downarrow w)$.
That is, for each child we individualize a different vertex of the target cell,
and then perform refinement on the partition.\footnote{An example of search tree with ordered partitions can be found in App.~\ref{app:tree}.}

We now define the canonical form of the abstract algorithm.
Note though that we will slightly alter this definition later in order to facilitate pruning of the search tree.
Recall that the associated ordered partition of each leaf $\tau$ is a discrete partition $\pi_\tau$,
which represents a candidate canonical ordering of the vertices.
Specifically, the permutation $\overline{\pi_\tau}$ maps the input vertices to their new index,
and the graph $G^{\overline{\pi_\tau}}$ is thus a candidate for the canonical form.
Let $L(G)$ denote all leaf nodes of the search tree starting from the graph $G$.
The canonical form $C(G)$ is then the permuted graph indicated by the leaf node with the representationally smallest permuted graph.
That is $C(G) = G^{\overline{\partition{\tau_{canon}}}}$ with $\tau_{canon} = \argmin_{\tau\in L(G)}\left\{G^{\overline{\partition{\tau}}}\right\}$.

\paragraph{Pruning with Automorphisms}
Let $\tau_a, \tau_b\in L(G)$ be distinct leaves of the search tree, for which the permuted graphs are representationally equal.
That is, with $\alpha = \overline{\partition{\tau_a}}$ and $\beta = \overline{\partition{\tau_b}}$ then $G^\alpha \repEq G^\beta$.
Permuting both sides with $\overline{\beta}$ gives $G^{\alpha\overline{\beta}} \repEq G^{\beta\overline{\beta}} = G$,
meaning that $\alpha\overline{\beta}$ is an automorphism of $G$.
During the tree traversal, when finding new leaves that give representations equal to our currently best leaf, we can derive an automorphism.
We call this an \emph{explicit automorphism}, and the complete automorphism group can be computed by considering all such pairs of leaf nodes \cite{pgi:2}.
Sometimes it is possible to deduce automorphisms from internal nodes of the search tree.
We call automorphisms found in this manner \emph{implicit automorphisms} \cite{pgi:1}.

During the canonization procedure, let $\tau = (v_1, v_2, \dots, v_k)$ be an internal node of the search tree and $u, v\in T(G, \pi_\tau)$ two vertices in the target cell for this node.
Further, let $\gamma\in \Aut(G)$ be some known automorphism that fixes all vertices individualized on the path from the root to $\tau$ but moves $u$ to $v$.
That is, $v_i^\gamma = v_i$ for $1\leq i\leq k$, and $u^\gamma = v$.
As $u$ and $v$ are equivalent under $\gamma$ the two subtrees rooted at $(v_1, v_2, \dots, v_k, u)$ and $(v_1, v_2, \dots, v_k, v)$ will be isomorphic, and we can safely skip traversal of one of them \cite[Operation $P_C$]{pgi:2}.

\paragraph{Pruning with Node Invariants}
During the construction of a tree node it is often possible to extract isomorphism invariant information.
The path from the root to a leaf thus has a sequence of such extracted information, which again is isomorphism invariant.
We then redefine the canonical form to be the one with the lexicographically smallest of such information sequences.

Formally, let $\mathcal{T}$ denote the set of all search tree nodes, and $\Omega$ an arbitrary totally ordered set.
An \emph{invariant function} $\phi\colon \mathcal{G}\times \mathcal{T}\rightarrow \Omega$ then assigns a value to each tree node.
The function must be isomorphism invariant, i.e., $\phi(G^\gamma, \tau^\gamma) = \phi(G, \tau)$.
For convenience we define the \emph{path invariant function} $\vec{\phi}\colon \mathcal{G}\times \mathcal{T}\rightarrow \Omega^*$.
For a tree node $\tau = (v_1, v_2, \dots, v_k)$ the value is $\vec{\phi}(G, \tau) = (\phi(G, v_1), \phi(G, v_2), \dots, \phi(G, v_k))$.
That is, it is the sequence of all node invariants from the root down to $\tau$.
We compare such sequences lexicographically.
Finally the canonical form is then redefined to be the one with the smallest permuted graph among the leaves with the smallest path invariant:
\begin{align*}
	\phi^* &= \min_{\tau\in L(G)} \vec{\phi}(\tau)																				\\
	\tau_{canon} &= \argmin_{\tau\in L(G)}\left\{G^{\overline{\partition{\tau}}}   \middle| \vec{\phi}(\tau) = \phi^*\right\}			\\
	C(G) &= G^{\overline{\partition{\tau_{canon}}}}
\end{align*}

\SetKwData{Visitor}{vis}

\section{A Generic Algorithm Framework}
\label{sec:framework}
From the abstract canonization algorithm we see that each concrete algorithm is defined by specific choices of sub-procedures for the six categories in the table below.
\begin{center}
\scriptsize
\begin{tabular}{@{}l@{}cc@{}}
\toprule
Extension Category					& Abstract Function	& Section	\\
\midrule
Tree traversal						& --- 																& \ref{sec:fw:tt}			\\
Target cell selection					& $T\colon \mathcal{G}\times \Pi\rightarrow 2^V$					& \ref{sec:fw:tc}			\\
Refinement							& $R\colon \mathcal{G}\times \Pi\rightarrow \Pi$						& \ref{sec:fw:r}			\\
Pruning with automorphisms			& ---																	& \ref{sec:fw:aut}		\\
Detection of implicit automorphisms	& ---																	& \ref{sec:fw:implicit}	\\
Node invariants						& $\phi\colon \mathcal{G}\times \mathcal{T}\rightarrow \Omega$		& \ref{sec:fw:inv}		\\
\bottomrule
\end{tabular}
\end{center}
The goal of the present framework is to provide an implementation of functionality common to all algorithms, and provide a facility for attaching extensions for the six categories.
Note that while only one option for target cell selection and tree traversal must be chosen, it can be beneficial to use multiple algorithms in conjunction for the remaining categories.

Using generic programming \cite{genericProgramming} we have designed a single common extension infrastructure,
based on the idea of a \emph{visitor}, similar to those used in the Boost Graph library \cite{GGCL,BGL}.
The core canonization procedure is given a visitor object \Visitor which must fulfill a collection of requirements,
i.e., it must model a specific \emph{Visitor} concept.
The concept specifies that a concrete visitor must implement a collection of callbacks that will be invoked at various points during algorithm.
Each visitor must additionally specify two data structures; one that will be instantiated per search tree, and one instantiated for each node in each search tree.%
\footnote{The details of the \emph{Visitor} concept can be found in App.~\ref{app:visitor}.}

We provide a compound visitor, which for a sequence of individual visitors aggregates the associated data structures and dispatches
method calls to all of the contained visitors.
The compound visitor enforces that exactly one visitor has implemented a tree traversal algorithm, and exactly one has a target cell selector.
For the following sections we assume the object \Visitor is such a compound visitor aggregating the sequence $(\Visitor_1, \Visitor_2, \dots, \Visitor_t)$ of visitors.

The common functionality of the framework further consists of data structures for tree nodes and ordered partitions,
along with methods for creation and destruction of tree nodes, including vertex individualization and invocation of target cell selection and refinement through the visitors.%
\footnote{Pseudocode for the framework methods can be found in the appendix, Fig.~\ref{fig:coreWrap}.}

In the following sections we provide further details for each of the six extension categories.
Outside those categories we provide two additional visitors; a debug visitor for collecting detailed logs,
and a stats visitor, e.g., for counting the number of tree nodes created and even for creating an annotated visualization of the search tree.%
\footnote{Examples of search tree visualizations can be found in App.~\ref{app:stats}.}%

\subsection{Tree Traversal and Automatic Garbage Collection}
\label{sec:fw:tt}
Most of the published algorithms and tools, including nauty \cite{pgi:1,pgi:2} and Bliss \cite{bliss,bliss:2} use depth-first traversal of the search tree.
Bliss notably exploits this traversal order to use a more efficient data structure for ordered partitions.
The tool Traces \cite{pgi:2} uses a different traversal scheme which combines a breadth-first traversal with so-called experimental paths to find leaves early.
As noted in \cite{cfi-rigid}, this means that Traces can consume much more memory than tools using depth-first traversal.

The framework directly allows for arbitrary tree traversal algorithms to be used, by defining appropriate visitors.
The lifetime of tree nodes is managed using reference counting,
where each node has a owning reference to its parent and the parent has a non-owning reference to its children.
Each visitor is thus responsible to keep owning references to nodes they are interested in.
The creation of new children is done through a framework method,
while discovered leaf nodes must be reported through another framework method that handles comparison of permuted graphs and potentially updating the current best leaf.
To facilitate pruning, a specific visitor method should be invoked before deciding which child node to create next.

We provide an implementation of the classical depth-first traversal (DFS)\footnote{Pseudocode for the implementation of DFS is shown in the appendix, Alg.~\ref{alg:dfs}.}
and a traversal similar to Traces (BFSExp).
We have also developed a memory sensitive hybrid of those two traversals (BFSExpM), which trades time for guaranteed memory usage.
Based on a given memory limit it uses BFSExp when the number of tree nodes is low, and uses DFS when above the limit.
It may therefore switch back to BFSExp if a sufficient amount of the search tree is deallocated.
With the provided debug visitor it is directly possible to investigate how
the number of currently allocated tree nodes develops through the course of the algorithm.%
\footnote{An example of investigating the number of tree nodes allocated can be found in App.~\ref{sec:app:maxNumNodes}.}%

\subsection{Target Cell Selection}
\label{sec:fw:tc}
A large variety of target cell selectors are available in Bliss, Traces, and nauty,
with the simplest selecting the first either smallest or largest cell.
A property used in more advanced target cell selectors is the following \cite{bliss}:
for two cells $U, W\in \pi$ we say that $U$ is \emph{non-uniformly joined} to $W$ if for all vertices $u\in U$
there are two vertices $w_1, w_2\in W$ such that $(u, v_1)\in E$ and $(u, v_2)\not\in E$.
That is, all vertices of $U$ have both neighbours and non-neighbours in $W$.

In the present framework the target cell selection is done during construction of each internal tree node,
using a dedicated visitor method. Exactly one visitor must indicate that it implemented this method.

We provide three target cell selectors: select the first non-singleton cell (F), select the first largest cell (FL),
and  select the first largest cell that is non-uniformly joined to the most cells (FLM).

\subsection{Refinement}
\label{sec:fw:r}
The basic refinement function used in most tools is the 1-dimensional Weisfeiler-Leman algorithm (WL-1)\cite{WL},
that iteratively separates vertices in a cell depending on their degree with respect to other cells.
Traces \cite{pgi:2} can also use the 2-dimensional variant.
The tool nauty includes a selection of additional refinement functions \cite{pgi:2}.

Refinement is invoked during the construction of a tree node, through the refinement visitor method.
Multiple visitors may do refinement, so the formal refinement function $R$ is derived from the composition $R_t\circ\dots\circ R_2\circ R_1$,
where $R_i$ is the refinement function implemented by visitor $\Visitor_i$.
The compound visitor coordinates the invocation using returned status codes, e.g., indicating whether any refinement was performed, or if it was aborted due to node invariant pruning.
To support calculation of node invariants and discovery of implicit automorphisms there are multiple visitor methods that refinement algorithms, especially WL-1, can call.
The simplest being the method called for each cell split performed.

We provide the WL-1 algorithm for refinement, which based on the observations in \cite{bliss}, uses custom implementations of counting sort and array partitioning to perform fast sorting for low degree cases.
As the framework directly allows for canonization of edge attributes, we have also generalized the WL-1 implementation to exploit the attributes for even stronger refinement.%
\footnote{The details of the generalized WL-1 algorithm can be found in App.~\ref{app:wl1}.}

\subsection{Pruning With Automorphisms}
\label{sec:fw:aut}
Let $A$ be the list of discovered automorphisms at some stage in the algorithm.
As automorphisms are closed under composition, we can consider the subgroup $\Phi\leq \Aut(G)$ generated from $A$.
For a tree node $\tau = (v_1, v_2, \dots, v_k)$ we are then interested in pruning with all permutations in the stabilizer of $\Phi$ with respect to the individualized vertices in $\tau$,
i.e., $\Stab_\Phi(\tau) = \{\gamma \in\Phi \mid v_i^\gamma = v_i, 1\leq i\leq k\}$.
This is for example done in Traces and newer versions of nauty \cite{pgi:2}, using random Schreier methods \cite{Seress:2003}.
A computational less intensive method is used in Bliss and earlier versions of nauty,
where only the subset $A_\tau =  \{\gamma \in A\mid v_i^\gamma = v_i, 1\leq
i\leq k\}$ is considered, i.e., a direct filtering of the found automorphisms without composition.
However, only a fixed number of permutations are stored at a time to
conserve memory \cite{bliss,pgi:1}.

The framework provides explicit automorphisms to the visitors, and
they may report implicit automorphisms to each other.
The actual pruning can be done in the visitor method that the tree traversal visitor is expected to invoke before deciding which child to create next.

We have implemented a visitor for automorphism pruning which itself is parameterized, such that it can work with different implementations of permutations, groups, and stabilizer chains.
For each tree node the visitor maintains the orbit partition of the target cell, using a union-find data structure.
In the present version we provide a parameterization that maintains $A_\tau$ in each node, similar to the strategy in Bliss and earlier versions of nauty
but without a limit on the number of stored automorphisms.\footnote{Pseudocode for the generic automorphism pruner visitor can be found in the appendix, Alg.~\ref{alg:autPruner}.}

\subsection{Detecting Implicit Automorphisms}
\label{sec:fw:implicit}
There are several known methods for finding additional automorphisms during the tree search.
For example, Lemma 2-25 \cite{pgi:1} describes three cases where all refinements of certain ordered partitions lead to isomorphic leaf nodes.
The simplest, and most common, of those cases is where the partition has singleton cells or cells of size 2.%
\footnote{Pseudocode for the simplest case of Lemma 2-25 \cite{pgi:1} can be found in Alg.~\ref{alg:invCell2}.}
Saucy \cite{saucy:1,saucy:2} introduced another heuristic for finding primarily sparse automorphisms, where for each non-leaf node a guess for an automorphism is made.
Traces \cite{pgi:2} reportedly generalizes this heuristic, though the details have not been described.
Traces also has heuristics for finding automorphisms when all vertices in non-singleton cells have certain degrees, but it is not clear what those heuristics are.

We have implemented two visitors in this category: one for the most common case of Lemma 2-25 \cite{pgi:1}, described above,
and one for refining cells with only degree 1 vertices and reporting the implicit automorphisms discovered in the process.

\subsection{Pruning With Node Invariants}
\label{sec:fw:inv}
In the abstract algorithm we used a single node invariant function, though in practice we may want to use multiple functions.
For example, one source of invariant data is from the sequence of cell splits performed by the refinement function, introduced in Traces \cite{pgi:2}.
Another important example is the \emph{partial leaf} invariant introduced in Bliss \cite{bliss}, calculated when new singleton cells arise in the refinement procedure.
Third, the WL-1 algorithm in general computes many different counts of edges, and this sequence of numbers can also be used as an invariant.
Importantly, both Bliss and nauty use hashing during node invariant computation, which may diminish the pruning power when collisions occur.

In the framework the calculation of and pruning with node invariants can be implemented entirely as visitors.
A visitor for coordinating multiple invariants is provided both for ensuring correct pruning, but also to minimize the overhead of implementing a new invariant.
Let $\Omega_i$ be the domain of invariant values produced by $\Visitor_i$,
then the node invariant values for a given tree node is a sequence of pairs $(\langle i_1, \omega_1\rangle, \langle i_2, \omega_2\rangle,\dots, \langle i_k, \omega_k\rangle)$
where $i_j$ is a visitor index and $\omega_j\in \Omega_{i_j}$.
The first component of each pair is stored in the coordinating visitor, while the second component is stored in the corresponding visitor.
The coordinating visitor handles invalidation of the current best leaf when a better path invariant is found,
and handles pruning of children of nodes that has already been created, but where a better invariant was found later.

We implement three invariants: the cell splitting trace introduced in Traces (T), a trace of values for the quotient graph also introduced in Traces \cite{pgi:2} (Q),
and the partial leaf invariant introduced in Bliss (PL).
Note that we do not hash the information in any of these visitors.

\section{Experimental Results}
\label{sec:results}
\newcommand\graphCol[1]{\texttt{#1}}
\newcommand\incPlot[5]{%
	\includegraphics[#5]{figures/#1/#2/#3/#4.pdf}%
}
\newcommand\incAll[4]{%
	\incPlot{All}{#1}{#2}{#3}{trim=0 2.3cm 0 0, clip, #4}%
}
\newcommand\incAllNoTrim[4]{%
	\incPlot{All}{#1}{#2}{#3}{trim=0 0.13cm 0 0, clip, #4}%
}
\newcommand\incAllPairFig[2]{%
	\subcaptionbox{\graphCol{#2}\label{fig:res:all:#2}}{%
		\incAll{#1}{#2}{stats}{width=0.52\textwidth}\hspace{-0.02\textwidth}%
		\incAll{#1}{#2}{benchmark}{width=0.52\textwidth}%
	}%
}
\newcommand\incAllTimeFig[3]{%
	\subcaptionbox{\graphCol{#2}\label{fig:res:all:#2}}{%
		\incAll{#1}{#2}{benchmark}{width=0.52\textwidth}\hspace{-0.02\textwidth}
		
		\vspace{-#3em}
	}%
}
\newcommand\incAllTimeFigNoTrim[3]{%
	\subcaptionbox{\graphCol{#2}\label{fig:res:all:#2}}{%
		\incAllNoTrim{#1}{#2}{benchmark}{width=0.52\textwidth}\hspace{-0.02\textwidth}
		
		\vspace{-#3em}
	}%
}

While the framework is capable of handling both vertex and edge attributes there are unfortunately no comprehensive collections with such graphs.
We have therefore benchmarked our implementation using both the collections of unattributed graphs from \cite{web:pipernoGraphs},
and using a proposed collection of difficult graphs, \graphCol{cfi-rigid} \cite{cfi-rigid,web:cfi-rigid}.
All executions were repeated 5 times with random permutations of the input graph, always with a maximum of \SI{8}{\giga\byte} memory and a 1000 second time limit.
Of the repetitions that succeeded we plot the average time spend, as well as markers if at least one execution ran out of memory (OOM) or out of time (OOT).
We have also recorded the number of tree nodes created, but as the elapsed time to a large degree is proportional to the number of nodes we only show time plots.
The full set of plots can be found at \cite{gh:graphCanon}. 
For comparison of absolute performance we also benchmarked the default configurations of Bliss v0.73 as well as v26r6 of nauty (dense and sparse) and Traces.
All experiments were run on compute nodes with two Intel E5-2680v3 CPUs (24 cores), using in total 12,000 compute node hours.
In the first part we investigate the effect of tree traversal algorithm and target cell selector,
while we focus in the second part on the \graphCol{cfi-rigid} collections where we also investigate the effect of different subsets of node invariants.
In the third part we illustrate how the BFSExpM traversal provides a memory-safe alternative to BFSExp at the expense of time.

\paragraph{Tree Traversal and Target Cell Selector}
For all graph collections from \cite{web:pipernoGraphs} we benchmarked the set $\{\mathrm{BFSExp},\mathrm{DFS}\}\times\{\mathrm{F},\mathrm{FL},\mathrm{FLM}\}$
of algorithm configurations, with all node invariants enabled.
Overall we found that no single configuration is the best performing on all instances, though BFSExp-FLM is often performing well.
In the following we focus on a selection of graph collections where we find that different subsets of algorithm configurations have significantly different
performance behaviors, and where some perform significantly better than the established tools.

On several of the (augmented) Miyazaki graphs we see that the performance largely is determined by the target cell selector.
Fig.~\ref{fig:res:all:mz-aug2} shows the result for the \graphCol{mz-aug2} collection,
\begin{figure*}
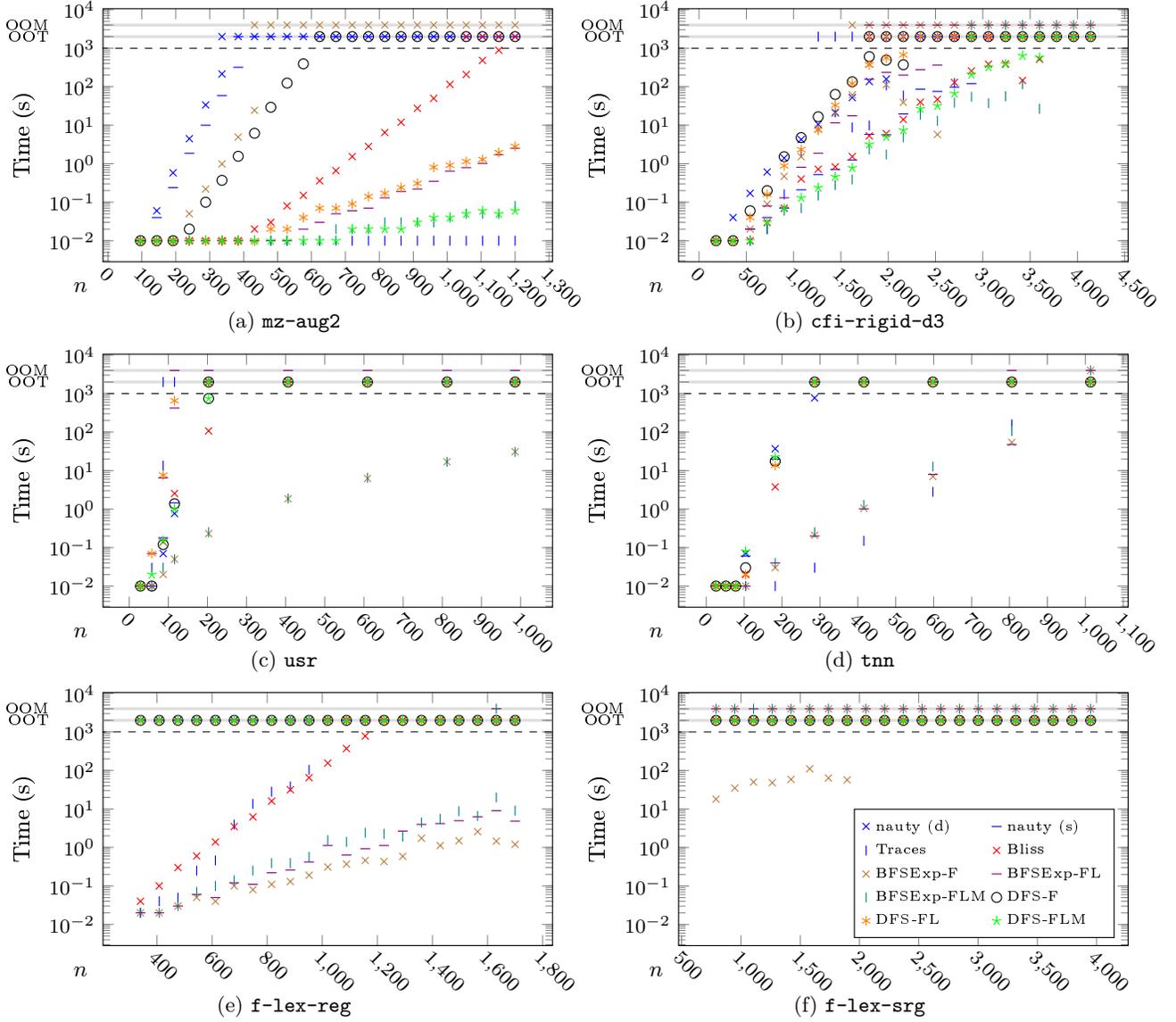

\centering
\incAllTimeFig{bliss}{mz-aug2}{1.7}%
\incAllTimeFig{cfi-rigid}{cfi-rigid-d3}{1.7}%

\incAllTimeFig{conauto}{usr}{2}%
\incAllTimeFig{conauto}{tnn}{2}%

\incAllTimeFig{misc}{f-lex-reg}{1.7}%
\incAllTimeFigNoTrim{misc}{f-lex-srg}{1.7}%

\caption[]{%
	Selected benchmark results for all combinations $\{\mathrm{BFSExp},\mathrm{DFS}\}\times\{\mathrm{F},\mathrm{FL},\mathrm{FLM}\}$, with all node invariants enabled.
}
\label{fig:res:all}
\end{figure*}
where the three pairs of F, FL, and FLM configurations have widely different performance.
From the number of tree nodes explored (see \cite{gh:graphCanon}) we see that the FLM target cell selector scales similar to Traces (sub-exponential),
while F and FL both have exponential behavior similar to Bliss and the two nauty modes.
This behavior we also see in the \graphCol{cmz} collection, while for \graphCol{mz-aug} only the F configurations scale exponentially.
In both \graphCol{mz-aug} and \graphCol{mz} the BFSExp-FLM configuration, for $n > 400$, will hit the memory limit for some executions but not all.
We attribute this to the automorphism pruning where we do not perform composition via random Schreier methods.
Thus, if the BFSExp tree traversal finds explicit automorphisms with few fixed vertices it may have fewer chances of using them for pruning.

On other collections, such as the non-disjoint union of tripartite graphs (\graphCol{tnn}, Fig.~\ref{fig:res:all:tnn}),
the algorithm configurations are separated by the tree traversal algorithms.
For \graphCol{tnn} the performance of the BFSExp configurations is similar to Traces (which is also breadth-first-based),
while for the union of strongly regular graphs (\graphCol{usr}, Fig.~\ref{fig:res:all:usr}) the BFSExp-F and BFSExp-FLM configurations perform distinctly better than all other algorithms.
Surprisingly the FL counterpart is one of the worst performing algorithms on the same collection.

The original collection of product graphs \graphCol{f-lex} contains two types of graphs where the algorithms perform differently, so we have split it into the two groups \graphCol{f-lex-reg} and \graphCol{f-lex-srg}, Fig.~\ref{fig:res:all:f-lex-reg} and \ref{fig:res:all:f-lex-srg}.
For \graphCol{f-lex-reg} we again see a separation by tree traversal algorithm, with the DFS configurations not being able to solve any instances.
However, Bliss also uses DFS but still solves many of the instances.
For the \graphCol{f-lex-srg} part of the collection only \graphCol{BFSExp-F} of all algorithms solve instances,
though only for some executions.

\paragraph{The \graphCol{cfi-rigid} Collections}
This package of six collections of graphs \cite{web:cfi-rigid},
was recently proposed \cite{cfi-rigid} explicitly as difficult instances for graph isomorphism,
and by construction they have very little symmetry.
Each collection (see Tab.~\ref{tab:cfi-rigid}) is constructed using a group, $D_3$, $\mathbb{Z}_3$, or $\mathbb{Z}_2$.
For $\mathbb{Z}_2$ there are further 3 variations where the instances
have gone through either a single or both of two reduction techniques (here denoted as $R^*$, $B^*$, and $R^*\circ B^*$).
\begin{table*}
\centering
\footnotesize
\begin{tabular}{@{}lcc@{\phantom{123456789}}lccr@{}}
\toprule
Col.				& Group 				& Reduction		& Best Algorithm		& Invariants Matter	& FLM Sep.	& Max.\ Solved $n$	\\
\midrule
\graphCol{d3}	& $D_3$				& ---				& BFSExp-FLM		& yes (any)			& yes	& 3,600	\\
\graphCol{z3}	& $\mathbb{Z}_3$	& ---				& BFSExp-FLM		& yes (any)			& yes	& 3,780	\\
\graphCol{z2}	& $\mathbb{Z}_2$	& ---				& Bliss, nauty (s)		& yes (any)			& no	& 2,992	\\
\graphCol{r2}	& $\mathbb{Z}_2$	& $R^*$			& Bliss, nauty (s)		& no				& no	& 1,584	\\
\graphCol{s2}	& $\mathbb{Z}_2$	& $B^*$			& FLM, Bliss, nauty (s)	& yes (PL or Q)		& no	& 2,496	\\
\graphCol{t2}	& $\mathbb{Z}_2$	& $R^*\circ B^*$	& FLM, Bliss, nauty (s)	& yes (PL or Q)		& yes	& 1,056	\\
\bottomrule
\end{tabular}
\caption[]{
	Summary of results for the \graphCol{cfi-rigid} collections.
	The right-most column is the largest instance that any of the configurations or the tools solved.
}
\label{tab:cfi-rigid}
\end{table*}
We have benchmarked all combinations, including subsets of node invariants, on all instances.
That is the 48 combinations
$\{\mathrm{BFSExp},\mathrm{DFS}\}\times\{\mathrm{F},\mathrm{FL},\mathrm{FLM}\}\times
2^{\{\mathrm{PL},\mathrm{Q},\mathrm{T}\}}$.

For the four $\mathbb{Z}_2$-based collections Bliss and nauty (sparse)
perform well, though both FLM configurations with all invariants have similar performance on \graphCol{s2} and \graphCol{t2}.
On the two other collections, \graphCol{d3} and \graphCol{z3}, the BFSExp-FLM configuration with all invariants is the best performing algorithm,
with the corresponding DFS-FLM configuration slightly behind.
There is however a significant separation up to F and FL configurations (Fig.~\ref{fig:res:all:cfi-rigid-d3}).
We do not see this separation in the plain $\mathbb{Z}_2$-based collection (\graphCol{z2}) or in those with just one reduction applied,
but interestingly the separation occurs when both reductions are applied at the same time, \graphCol{t2}.

In the investigation of the effect of node invariants we first of all found that the relative effect on performance is independent of tree traversal and target cell selector, within the same graph collection.
For \graphCol{d3}, \graphCol{z3}, and \graphCol{z2}, i.e., the collections without reductions, it improves performance when enabling any one invariant.
Though the effect of enabling additional invariants is minor or non-existing.
Intriguingly, for \graphCol{r2} where the $R^*$ reduction were used to create the instances, the node invariants do not seem to have any effect at all.
When the other reduction, $B^*$, has been applied (\graphCol{s2}), we find that the cell splitting invariant (T) has no effect,
but either PL, Q, or both have the same improving effect.
Surprisingly, we also see that pattern of effect on \graphCol{t2} which has undergone both reductions.
We hope that future in-depth studies on these graphs and node invariants may lead to new insights, both for developing better invariants
but potentially also for creating even more difficult benchmark graphs.

\begin{figure}
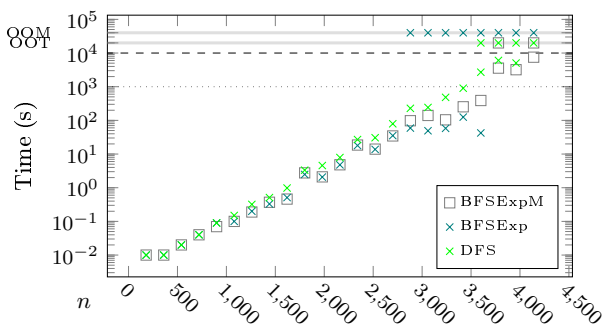

\centering
\incPlot{Mem}{cfi-rigid}{cfi-rigid-d3}{benchmark}{trim=0 0.1cm 0em 0, width=0.48\textwidth}
\caption[]{
Comparison of tree traversal algorithms on \graphCol{cfi-rigid-d3}, using FLM for target cell selection and all three node invariants.
}
\label{fig:res:mem:cfi-rigid-d3}
\end{figure}

\paragraph{Memory Sensitive BFSExp}
In some contexts it is highly undesirable to run out of memory, and we have therefore developed the BFSExpM tree traversal as described earlier.
We have tested it using FLM as target cell selector on \graphCol{cfi-rigid-d3} with a limit to ensure the whole process does not hit the \SI{8}{\giga\byte} hard-cap.
Varying the memory limit (down to \SI{1}{\giga\byte}) did not result in different performances.
In Fig.~\ref{fig:res:mem:cfi-rigid-d3} a comparison is shown with the BFSExp and DFS configurations.
We clearly see that for the instance sizes where BFSExp go out of memory on some executions,
the BFSExpM configuration increase in average time spend.
Notably it still performs better than DFS, thereby being a viable alternative when a memory limit must be honored.
For this experiment we let all executions run for 10,000 seconds, and we see that BFSExpM is the only configuration to solve the largest instance, though with some executions exceeding the time limit.

\section{Concluding Remarks}
\label{sec:conclusion}

We have presented a versatile framework for fast graph canonization algorithms that makes it easy to implement and compare heuristics.
Not only does it perform better than the established tools on several graph classes,
but we find interesting performance separations between different choices of heuristics which is not immediately possible with the established tools.
In the future we will expand the set of available heuristics, to approach a full algorithm library for graph canonization.
While the established tools can handle graphs with vertex attributes, the presented framework can also directly handle edge attributes, and even exploit them for refinement.
Though, the attribute sets are currently limited to totally ordered sets, which is not general enough to for example handle attributes used for encoding stereo-chemistry in molecule graphs \cite{stereoGraTra,Heller2015}.
In our preliminary investigations we find that it is possible to lift this restriction, and that a novel type of node invariant can be introduced to exploit complex attributes for pruning.

\section*{Acknowledgements}
This work was supported in part by the COST-Action CM1304 ``Systems Chemistry'',
by the Independent Research Fund Denmark, Natural Sciences, grant DFF-7014-00041,
and by the ELSI Origins Network (EON), which is supported by a grant from the
John Templeton Foundation. Computation for the work described in this paper was
supported by the DeiC National HPC Centre, SDU.
The opinions expressed in this publication are those of the authors and do not
necessarily reflect the views of the John Templeton Foundation.

\bibliography{paper}

\appendix
\section{Comparison of Graph Representations}
\label{app:adjRep}

The canonization algorithm relies on a total order to exist on the set of all graph representations.
For adjacency matrices we can find the order among two graphs for example by lexicographic comparison of the matrices.
When edge attributes are present we can imagine them being stored in the matrix and then require the attribute domain to be totally ordered.
With vertex attributes we can simply modify the comparison such that we first lexicographically compare vertex attributes in order of the vertex index, and then the matrices.

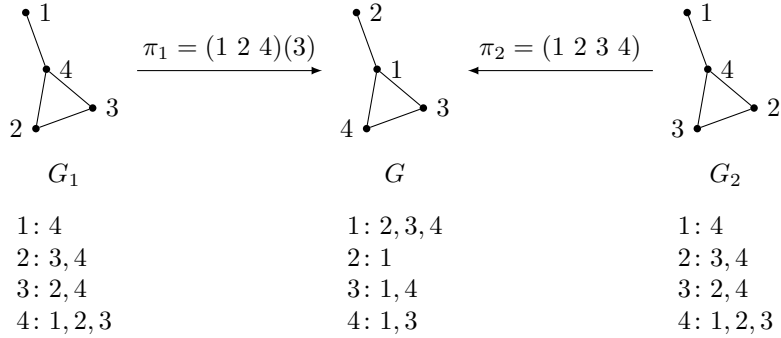
\begin{figure*}
\centering
\begin{tikzpicture}[node distance=7em]
\node[label=below:$G_1$] (G1) {
\begin{tikzpicture}[gDot, node distance=2em]
\node[vDot, label=right:1] (v-1) {};
\node[vDot, label=right:4, position=-70 degrees from v-1] (v-2) {};
\node[vDot, label=right:3, position=-40 degrees from v-2] (v-3) {};
\node[vDot, label=left:2, position=-100 degrees from v-2] (v-4) {};
\foreach \s/\t in {1/2, 2/3, 2/4, 3/4}
	\draw[eDot] (v-\s) to (v-\t);
\end{tikzpicture}
};
\node[label=below:$G$, right=of G1] (G) {
\begin{tikzpicture}[gDot, node distance=2em]
\node[vDot, label=right:2] (v-1) {};
\node[vDot, label=right:1, position=-70 degrees from v-1] (v-2) {};
\node[vDot, label=right:3, position=-40 degrees from v-2] (v-3) {};
\node[vDot, label=left:4, position=-100 degrees from v-2] (v-4) {};
\foreach \s/\t in {1/2, 2/3, 2/4, 3/4}
	\draw[eDot] (v-\s) to (v-\t);
\end{tikzpicture}
};
\node[label=below:$G_2$, right=of G] (G2) {
\begin{tikzpicture}[gDot, node distance=2em]
\node[vDot, label=right:1] (v-1) {};
\node[vDot, label=right:4, position=-70 degrees from v-1] (v-2) {};
\node[vDot, label=right:2, position=-40 degrees from v-2] (v-3) {};
\node[vDot, label=left:3, position=-100 degrees from v-2] (v-4) {};
\foreach \s/\t in {1/2, 2/3, 2/4, 3/4}
	\draw[eDot] (v-\s) to (v-\t);
\end{tikzpicture}
};
\draw[->, >=latex] (G1) to node[above] {$\pi_1 = (1\ 2\ 4)(3)$} (G);
\draw[->, >=latex] (G2) to node[above] {$\pi_2 = (1\ 2\ 3\ 4)$} (G);

\node[below=2em of G1, align=left] {
$1\colon 4$\\
$2\colon 3, 4$\\
$3\colon 2, 4$\\
$4\colon 1, 2, 3$
};
\node[below=2em of G, align=left] {
$1\colon 2, 3, 4$\\
$2\colon 1$\\
$3\colon 1, 4$\\
$4\colon 1, 3$
};
\node[below=2em of G2, align=left] {
$1\colon 4$\\
$2\colon 3, 4$\\
$3\colon 2, 4$\\
$4\colon 1, 2, 3$
};
\end{tikzpicture}
\caption[Permutation of graphs]{
Three isomorphic graphs represented as adjacency lists.
The underlying indices of the vertices are shown,
and the permutations $\pi_1$ and $\pi_2$ (in cycle notation) describe the relationship between the indices of the graphs.
From the adjacency lists we see that $G_1$ is representationally equal to $G_2$, and representationally different from $G$.
The permutation $\pi' = \pi_1\overline{\pi_2} = (2\ 3)(1)(4)$ thus represents an isomorphism from $G_1$ to $G_2$.
}
\label{fig:canon:exSmall}
\end{figure*}

For adjacency lists a similar comparison can be defined. Assuming the vertex indices are defined by the position of the vertices in the data structure,
we still have freedom to order each of the lists of incident edges.
We can say that an adjacency list is \emph{globally ordered} if each edge list is sorted by the neighboring vertex index,
and with edge attributes and multigraphs we further require parallel edges to be ordered by the edge attribute.
Two globally ordered adjacency lists can then be compared lexicographically in the natural manner.
Vertex attributes can be trivially incorporated in this procedure.
An example of globally ordered adjacency lists are shown in  Fig.~\ref{fig:canon:exSmall},
along with a visualization of how automorphisms can be detected by comparing graph representations from different permutations of the input indices.

\tikzset{gCanon/.style={scale=0.75}}
\tikzset{vCanon/.style={draw, circle, inner sep=0, minimum size=1.3em, font=\scriptsize}}
\newcommand\canonPi[1]{\begingroup\scriptsize #1\endgroup}
\tikzset{canon-c-1/.style={text=black, fill=white}}
\tikzset{canon-c-2/.style={text=black, fill=green}}
\tikzset{canon-c-3/.style={text=white, fill=ForestGreen}}
\tikzset{canon-c-4/.style={text=white, fill=blue}}
\tikzset{canon-c-5/.style={text=black, fill=cyan}}
\tikzset{canon-c-6/.style={text=black, fill=orange}}
\tikzset{canon-c-7/.style={text=white, fill=red}}
\tikzset{canon-c-8/.style={text=white, fill=DarkOrchid}}
\tikzset{canon-c-9/.style={text=black, fill=gray}}
\tikzset{canon-c-10/.style={text=white, fill=black}}
\newcommand\canonCoords{
\coordinate (v-1-c) at (-1.5418, -0.83072) {};
\coordinate (v-2-c) at (1.439, 0.77856) {};
\coordinate (v-3-c) at (-0.50097, -0.85872) {};
\coordinate (v-4-c) at (-0.98948, 0.049983) {};
\coordinate (v-5-c) at (0.38833, 1.319) {};
\coordinate (v-6-c) at (1.3187, -0.39764) {};
\coordinate (v-7-c) at (-0.22862, 0.47222) {};
\coordinate (v-8-c) at (0.27362, -0.44751) {};
\coordinate (v-9-c) at (-0.73546, 1.1707) {};
\coordinate (v-10-c) at (0.57657, -1.2558) {};
}
\newcommand\canonEdges{
\foreach \s/\t in {1/3, 1/4, 2/5, 2/6, 3/7, 3/10, 4/8, 4/9, 5/7, 5/9, 6/8, 6/10, 7/9, 8/10}
	\draw (v-\s) to (v-\t);
}

\section{Search Tree Example}
\label{app:tree}
\begin{figure*}
\centering
\newcommand{\canonSpace}{2em}
\tikzset{gWrap/.style={align=center, inner sep=0}}

\begin{tikzpicture}[gCanon, remember picture]
\node[gWrap] (g-1) {
\begin{tikzpicture}
\canonCoords
\foreach \i in {1} \node[vCanon, canon-c-1] (v-\i) at (v-\i-c) {\i};
\foreach \i in {2} \node[vCanon, canon-c-2] (v-\i) at (v-\i-c) {\i};
\foreach \i in {7, ..., 10}
	\node[vCanon, canon-c-3] (v-\i) at (v-\i-c) {\i};
\foreach \i in {5, 6}
	\node[vCanon, canon-c-7] (v-\i) at (v-\i-c) {\i};
\foreach \i in {3, 4}
	\node[vCanon, canon-c-9] (v-\i) at (v-\i-c) {\i};
\canonEdges
\end{tikzpicture}\\
\canonPi{$\pi_{(1)} = [1\mid 2\mid 7\ 8\ 9\ 10\mid 5\ 6\mid 3\ 4]$}
};
\end{tikzpicture}
\hfill
\begin{tikzpicture}[gCanon, remember picture]
\node[gWrap] (g) {
\begin{tikzpicture}
\canonCoords
\foreach \i in {1, 2}
	\node[vCanon, canon-c-1] (v-\i) at (v-\i-c) {\i};
\foreach \i in {7, ..., 10}
	\node[vCanon, canon-c-3] (v-\i) at (v-\i-c) {\i};
\foreach \i in {3, ..., 6}
	\node[vCanon, canon-c-7] (v-\i) at (v-\i-c) {\i};
\canonEdges
\end{tikzpicture}\\
\canonPi{$\pi_{()} = [1\ 2\mid 7\ 8\ 9\ 10\mid 3\ 4\ 5\ 6]$}
};
\end{tikzpicture}
\hfill
\begin{tikzpicture}[gCanon, remember picture]
\node[gWrap] (g-2) {
\begin{tikzpicture}
\canonCoords
\foreach \i in {2} \node[vCanon, canon-c-1] (v-\i) at (v-\i-c) {\i};
\foreach \i in {1} \node[vCanon, canon-c-2] (v-\i) at (v-\i-c) {\i};
\foreach \i in {7, ..., 10}
	\node[vCanon, canon-c-3] (v-\i) at (v-\i-c) {\i};
\foreach \i in {3, 4}
	\node[vCanon, canon-c-7] (v-\i) at (v-\i-c) {\i};
\foreach \i in {5, 6}
	\node[vCanon, canon-c-9] (v-\i) at (v-\i-c) {\i};
\canonEdges
\end{tikzpicture}\\
\canonPi{$\pi_{(2)} = [2\mid 1\mid 7\ 8\ 9\ 10\mid 3\ 4\mid 5\ 6]$}
};
\end{tikzpicture}\\[\canonSpace]

\begin{tikzpicture}[gCanon, remember picture]
\node[gWrap, align=left] (g-1-7) {
\begin{tikzpicture}
\canonCoords
\foreach \i in {1} \node[vCanon, canon-c-1] (v-\i) at (v-\i-c) {\i};
\foreach \i in {2} \node[vCanon, canon-c-2] (v-\i) at (v-\i-c) {\i};
\foreach \i in {7}	\node[vCanon, canon-c-3] (v-\i) at (v-\i-c) {\i};
\foreach \i in {10} \node[vCanon, canon-c-4] (v-\i) at (v-\i-c) {\i};
\foreach \i in {8} \node[vCanon, canon-c-5] (v-\i) at (v-\i-c) {\i};
\foreach \i in {9} \node[vCanon, canon-c-6] (v-\i) at (v-\i-c) {\i};
\foreach \i in {6} \node[vCanon, canon-c-7] (v-\i) at (v-\i-c) {\i};
\foreach \i in {5} \node[vCanon, canon-c-8] (v-\i) at (v-\i-c) {\i};
\foreach \i in {4}	\node[vCanon, canon-c-9] (v-\i) at (v-\i-c) {\i};
\foreach \i in {3} \node[vCanon, canon-c-10] (v-\i) at (v-\i-c) {\i};
\canonEdges
\end{tikzpicture}\\
\canonPi{$\pi_{(1, 7)} = [1\mid 2\mid 7\mid 10\mid 8\mid 9\mid 6\mid 5\mid 4\mid 3]$}
};
\end{tikzpicture}
\hfill
\hspace{-5em}
\begin{tikzpicture}[remember picture,
	vTree/.style={draw, circle, minimum size=1.3em, inner sep=0, font=\scriptsize},
	pruned/.style={gray},
	level 1/.style={sibling distance=6em},
	level 2/.style={sibling distance=1.5em}
]
\node[vTree] (t) {}
	child {node[vTree] (t-1) {1}
		child {node[vTree] (t-1-7) {7}}
		child {node[vTree] (t-1-8) {8}}
		child {node[vTree] (t-1-9) {9}}
		child {node[vTree, pruned] (t-1-10) {10}}
	}
	child {node[vTree] (t-2) {2}
		child {node[vTree] (t-2-7) {7}}
		child {node[vTree, pruned] (t-2-8) {8}}
		child {node[vTree, pruned] (t-2-9) {9}}
		child {node[vTree, pruned] (t-2-10) {10}}
	}
;
\node[at=(t-1-10.south), minimum size=3em] {};
\end{tikzpicture}
\hspace{-5em}
\hfill
\begin{tikzpicture}[gCanon, remember picture]
\node[gWrap, align=right] (g-2-7) {
\begin{tikzpicture}
\canonCoords
\foreach \i in {2} \node[vCanon, canon-c-1] (v-\i) at (v-\i-c) {\i};
\foreach \i in {1} \node[vCanon, canon-c-2] (v-\i) at (v-\i-c) {\i};
\foreach \i in {7} \node[vCanon, canon-c-3] (v-\i) at (v-\i-c) {\i};

\foreach \i in {10} \node[vCanon, canon-c-4] (v-\i) at (v-\i-c) {\i};
\foreach \i in {8} \node[vCanon, canon-c-5] (v-\i) at (v-\i-c) {\i};

\foreach \i in {9} \node[vCanon, canon-c-6] (v-\i) at (v-\i-c) {\i};

\foreach \i in {4} \node[vCanon, canon-c-7] (v-\i) at (v-\i-c) {\i};
\foreach \i in {3} \node[vCanon, canon-c-8] (v-\i) at (v-\i-c) {\i};

\foreach \i in {6} \node[vCanon, canon-c-9] (v-\i) at (v-\i-c) {\i};
\foreach \i in {5} \node[vCanon, canon-c-10] (v-\i) at (v-\i-c) {\i};
\canonEdges
\end{tikzpicture}\\
\canonPi{$\pi_{(2,7)} = [2\mid 1\mid 7\mid 10\mid 8\mid 9\mid 4\mid 3\mid 6\mid 5]$}
};
\end{tikzpicture}\\[\canonSpace]

\begin{tikzpicture}[gCanon, remember picture]
\node[gWrap] (g-1-8) {
\begin{tikzpicture}
\canonCoords
\foreach \i in {1} \node[vCanon, canon-c-1] (v-\i) at (v-\i-c) {\i};
\foreach \i in {2} \node[vCanon, canon-c-2] (v-\i) at (v-\i-c) {\i};
\foreach \i in {8} \node[vCanon, canon-c-3] (v-\i) at (v-\i-c) {\i};
\foreach \i in {9} \node[vCanon, canon-c-4] (v-\i) at (v-\i-c) {\i};
\foreach \i in {7} \node[vCanon, canon-c-5] (v-\i) at (v-\i-c) {\i};
\foreach \i in {10} \node[vCanon, canon-c-6] (v-\i) at (v-\i-c) {\i};
\foreach \i in {5} \node[vCanon, canon-c-7] (v-\i) at (v-\i-c) {\i};
\foreach \i in {6} \node[vCanon, canon-c-8] (v-\i) at (v-\i-c) {\i};
\foreach \i in {3} \node[vCanon, canon-c-9] (v-\i) at (v-\i-c) {\i};
\foreach \i in {4} \node[vCanon, canon-c-10] (v-\i) at (v-\i-c) {\i};
\canonEdges
\end{tikzpicture}\\
\canonPi{$\pi_{(1, 8)} = [1\mid 2\mid 8\mid 9\mid 7\mid 10\mid 5\mid 6\mid 3\mid 4]$}
};
\end{tikzpicture}
\hfill
\begin{tikzpicture}
\foreach \i in {1, ..., 10} {
	\node[draw, canon-c-\i, minimum width=1.8em, minimum height=1em, inner sep=0] at (0, \i*-1em) (c-\i) {\scriptsize\i};
};
\node[anchor=north, at=(c-10.south)] {\scriptsize Colour order};
\end{tikzpicture}
\hfill
\begin{tikzpicture}[gCanon, remember picture]
\node[gWrap] (g-1-9) {
\begin{tikzpicture}
\canonCoords
\foreach \i in {1} \node[vCanon, canon-c-1] (v-\i) at (v-\i-c) {\i};
\foreach \i in {2} \node[vCanon, canon-c-2] (v-\i) at (v-\i-c) {\i};
\foreach \i in {9} \node[vCanon, canon-c-3] (v-\i) at (v-\i-c) {\i};

\foreach \i in {8} \node[vCanon, canon-c-4] (v-\i) at (v-\i-c) {\i};
\foreach \i in {10} \node[vCanon, canon-c-5] (v-\i) at (v-\i-c) {\i};

\foreach \i in {7} \node[vCanon, canon-c-6] (v-\i) at (v-\i-c) {\i};
\foreach \i in {6} \node[vCanon, canon-c-7] (v-\i) at (v-\i-c) {\i};
\foreach \i in {5} \node[vCanon, canon-c-8] (v-\i) at (v-\i-c) {\i};

\foreach \i in {3} \node[vCanon, canon-c-9] (v-\i) at (v-\i-c) {\i};
\foreach \i in {4} \node[vCanon, canon-c-10] (v-\i) at (v-\i-c) {\i};
\canonEdges
\end{tikzpicture}\\
\canonPi{$\pi_{(1, 9)} = [1\mid 2\mid 9\mid 8\mid 10\mid 7\mid 6\mid 5\mid 3\mid 4]$}
};
\end{tikzpicture}
\begin{tikzpicture}[remember picture, overlay,
	e/.style={dashed}
]
\draw[e] (g) to (t);
\draw[e] (g-1) to[bend right=10] (t-1);
\draw[e] (g-2) to[bend left=10] (t-2);
\draw[e] (g-1-7) ++(3em, -2em) to (t-1-7);
\draw[e] (g-1-8) ++(5em, 4.5em) to[in=-45, out=45] (t-1-8);
\draw[e] (g-1-9) ++(-4.5em, 5em) to[in=-35, out=150] (t-1-9);
\draw[e] (g-2-7) ++(-6.8em, -4.2em) to[in=-45, out=-180] (t-2-7);
\end{tikzpicture}
\caption{
A search tree starting with the refinement of the unit partition in the root.
The refinement function used is the WL-1 algorithm, the target cell selector is selecting the first non-singleton cell,
and no node invariants are used.
Each node in the tree represents a sequence of vertex individualizations, where the latest vertex being individualized is shown in the nodes.
For most tree nodes the corresponding partition is shown along with the colored graph it represents.
In the colored graphs the vertices are labeled with the vertex indices from the input graph,
and colored with ``numbers'' corresponding to the potential canonical vertex indices.
Note that the colored graphs in the leaves of the left half of the tree (the children of $\tau = (1)$) are all isomorphic.
This is also true among the children in the right half of the tree (the children of $\tau = (2)$.
However between the halves of the tree, the graphs are not isomorphic.
The grayed out nodes correspond to nodes pruned from automorphism discovery, when depth-first traversal of the tree is used.
The example is heavily inspired by \cite[Figure 3]{piperno}, though here using different functions for refinement and target cell selection.
}
\label{fig:canon:tree}
\end{figure*}
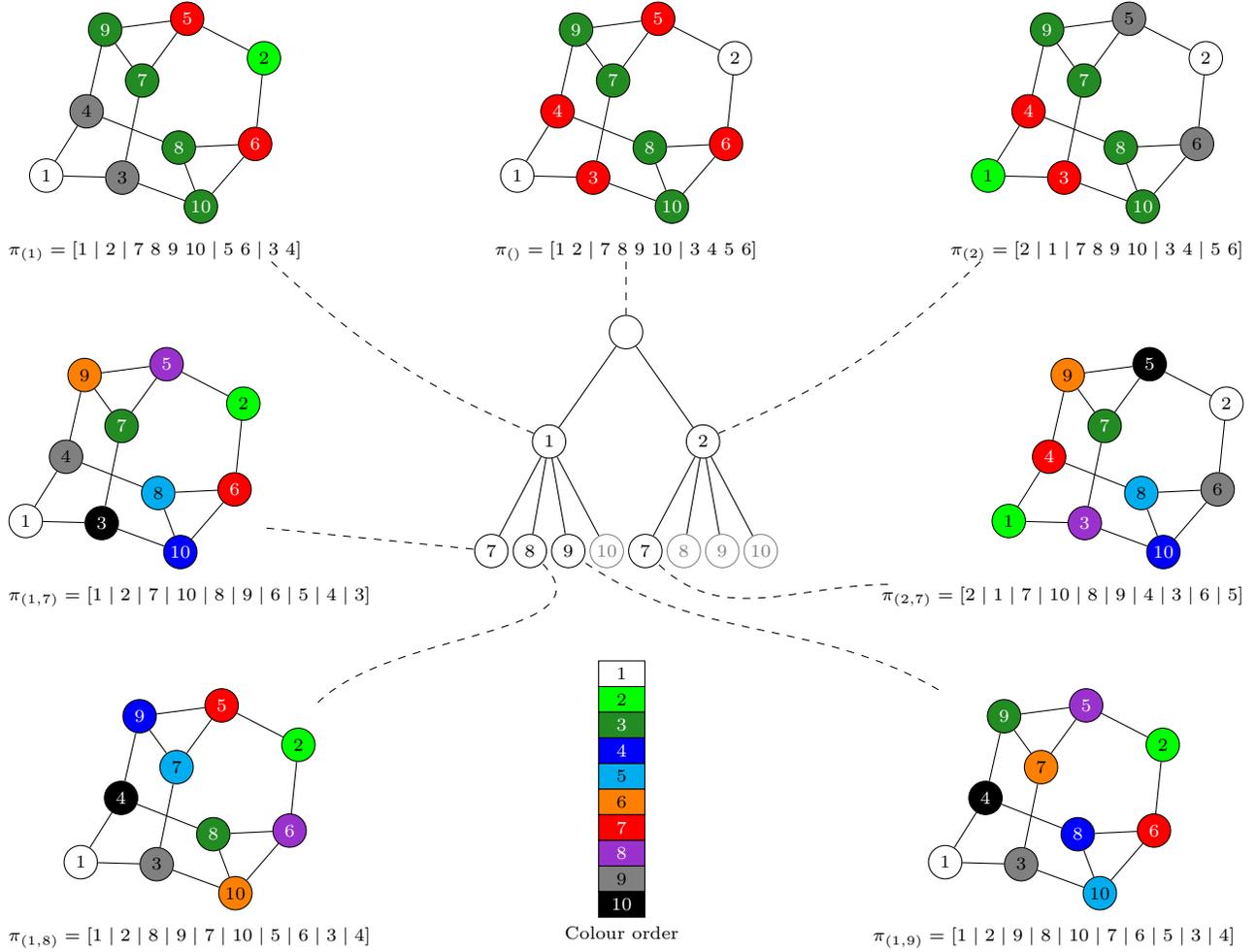

An example of a search tree is shown in Fig.~\ref{fig:canon:tree}.

\newcommand\refAlgLine[2]{Alg.~\ref{alg:#1}, line~\ref{alg:line:#2}}
\newcommand\assign{\ensuremath{\leftarrow}\ }
\newcommand\inlineCode[1]{\texttt{#1}}

\newcommand\algTextSize\scriptsize
\newcommand\algwrapRatio{0.49}
\SetInd{0.5em}{0em}
\SetAlgoHangIndent{1em}
\DecMargin{0.5ex}

\SetKwProg{VisFn}{@}{\string:}{}

\newcommand\codeLinkText{$\vartriangleright$C++}
\newcommand\codeLink[2]{%
	\href{https://github.com/jakobandersen/graph_canon/blob/v0.1/include/graph_canon/#1\#L#2}{\codeLinkText}%
}
\newcommand\DefMethod[4]{%
	\SetKwFunction{#1}{#2}%
	\SetKwProg{Fn#1}{def}{\string:\hfill\codeLink{#3}{#4}}{}
}
\SetKwProg{Fn}{def}{\string:}{}
\SetKwFunction{LCA}{lca}
\DefMethod{Canonicalize}{canon}{canonicalization.hpp}{281}
\SetKwData{VComp}{vComp}
\SetKwData{EHandler}{eHandler}
\SetKwFunction{Pos}{pos}
\SetKwFunction{Vertex}{vertex}
\SetKwFunction{NonOwningRef}{nonOwningRef}
\DefMethod{MakeTreeNode}{makeTreeNode}{detail/tree_node.hpp}{25}
\DefMethod{MakeChild}{makeChildNode}{detail/tree_node.hpp}{87}
\DefMethod{PruneTree}{pruneTree}{detail/tree_node.hpp}{122}
\DefMethod{DestroyTreeNode}{destroyTreeNode}{detail/tree_node.hpp}{50}
\SetKwArray{Child}{child}
\SetKwArray{ChildPruned}{childPruned}
\SetKwData{Parent}{parent}
\SetKwData{IsPruned}{isPruned}
\SetKwData{TargetCell}{targetCell}
\SetKwData{IndVertex}{individualizedVertex}
\SetKwData{Stab}{stab}	
\SetKwData{Fulfilled}{fulfilled} 
\DefMethod{AddTerminal}{addLeaf}{canonicalization.hpp}{218}
\SetKwData{VisitorType}{Vis}
\SetKwData{TreeDataType}{TreeData}
\SetKwData{NodeDataType}{NodeData}
\SetKwData{CanSelectTargetCellType}{CanSelectTargetCell}
\SetKwData{CanTraverseTreeType}{CanTraverseTree}
\SetKwData{TrueType}{TrueType}
\SetKwData{FalseType}{FalseType}
\SetKwData{Position}{position}

\SetKwData{CanonLeaf}{canonLeaf}
\SetKwInput{InstanceData}{Tree data}
\SetKwInput{TreeNodeData}{Node data}
\SetKwFunction{Traverse}{traverseTree}
\SetKwFunction{SelectTargetCell}{selectTargetCell}
\SetKwFunction{TreeNodeBegin}{treeNodeCreateBegin}
\SetKwFunction{TreeNodeEnd}{treeNodeCreateEnd}
\SetKwFunction{TreeNodeDestroy}{treeNodeDestroy}
\SetKwFunction{Refine}{refine}
\SetKwFunction{RefineAbort}{refineAbort}
\SetKwFunction{BeforeDescend}{beforeDescend}
\SetKwFunction{IsomorphicLeaf}{isomorphicLeaf}
\SetKwFunction{ImplicitAutomorphism}{implicitAutomorphism}
\SetKwFunction{NewCell}{newCell}
\SetKwData{Generation}{generation}
\SetKwData{CurGeneration}{generation}
\SetKwData{PathInv}{pathInv}
\SetKwData{PathGen}{pathGen}
\SetKwFunction{StackPush}{push}
\SetKwFunction{StackPop}{pop}
\SetKwFunction{FifoPush}{pushBack}
\SetKwFunction{FifoPop}{popFront}
\SetKwFunction{FifoRemove}{remove}
\SetKwInput{Input}{Input}
\SetKwInput{Data}{Data}
\SetKwArray{EdgeCount}{$C$}

\section{Framework Details}
\label{app:code}

\subsection{Visitor Concept}
\label{app:visitor}

The following is an outline of the \emph{Visitor} concept, ommitting technical details and several methods, e.g., for callbacks during refinement.
The concept is not explicitly codified in the implementation, but the compound visitor (\codeLink{visitor/compound.hpp}{54}) in practice enforces it.

A type \VisitorType satisfies the \emph{Visitor} concept if the following requirements are fulfilled.

\subsubsection*{Associated Types}
\VisitorType must have the following nested types:
\begin{itemize}
\item \TreeDataType: the type of a data structure to be instantiated for each search tree.
\item \NodeDataType: the type of a data structure to be instantiated for each search tree node.
\item \CanSelectTargetCellType: a type convertible to \TrueType or \FalseType indicating whether the visitor implements target cell selection.
\item \CanTraverseTreeType: a type convertible to \TrueType or \FalseType indicating whether the visitor implements tree traversal.
\end{itemize}

\subsubsection*{Syntax}
\begin{itemize}
\item \Visitor, an object of type \VisitorType
\item $\tau_{root}$, a tree node representing the root of a search tree
\item $\tau$, an arbitrary tree node
\item $\gamma$, a non-trivial permutation in $S_n$
\item $\Position$, an integer indicating the start of a cell
\end{itemize}

\subsubsection*{Valid Expressions}
\begin{itemize}
\item \Visitor.\Traverse{$\tau_{root}$}, if \CanSelectTargetCellType is convertible to \TrueType\\
	Must implement a tree traversal algorithm.
	Called from \Canonicalize, \refAlgLine{core}{traverse}.
\item \Visitor.\SelectTargetCell{$\tau$}, if \CanSelectTargetCellType is convertible to \TrueType\\
	Must implement a target cell selector, $T$.
	Called from \MakeTreeNode, \refAlgLine{tree}{targetCell}.
\item \Visitor.\IsomorphicLeaf{$\tau$}\\
	Called from \AddTerminal, \refAlgLine{tt}{isoLeaf}.
\item \Visitor.\ImplicitAutomorphism{$\gamma$}\\
	May be called at any time by visitors.
\item \Visitor.\TreeNodeBegin{$\tau$}\\
	Called from \MakeTreeNode, \refAlgLine{tree}{treeBegin}.
\item \Visitor.\TreeNodeEnd{$\tau$}\\
	Called from \MakeTreeNode, \refAlgLine{tree}{treeEnd}.
\item \Visitor.\TreeNodeDestroy{$\tau$}
	Called from \DestroyTreeNode, \refAlgLine{tree}{treeDestroy}.
\item \Visitor.\Refine{$\tau$}\\
	Must implement a refinement function $R$, but in-place.
	Must call \RefineAbort on the overall visitor object if it returns due to the tree node becomming pruned.
	Called from \MakeTreeNode, \refAlgLine{tree}{refine}.
\item \Visitor.\RefineAbort{$\tau$}\\
	Called by refinement functions.
\item \Visitor.\BeforeDescend{$\tau$}\\
	Should be called by tree traversal algorithms before deciding which child to create next.
\item \Visitor.\NewCell{$\tau$, $position$}\\
	Must be called by refinement functions for each new cell split.
\end{itemize}

\subsection{Pseudocode for Framework Methods}
The pseudocode is shown in Fig.~\ref{fig:coreWrap}.
\begin{figure*}[tbp]
\begin{minipage}[t][36em]{\algwrapRatio\textwidth}
\removelatexerror
\begin{algorithm2e}[H]\algTextSize
\caption{Canonization Function}\label{alg:core}
\FnCanonicalize{\Canonicalize{$G$, \Visitor, \VComp, \EHandler}} {
	\tcp{We implicitly assume that references to\\ the following variables are passed \\recursively to all functions:\\$G$, \Visitor, \EHandler, and \CanonLeaf.}
	\CanonLeaf			\assign\Nil\;
	$\pi_0$	\assign \emph{The ordered partition $(V_1, V_2, \dots, V_k)$
		as described in Sec.~\ref{sec:abstractAlg}, but using \VComp for vertex comparison. \codeLink{canonicalization.hpp}{173}}\;
	$\tau_{root}$							\assign \MakeTreeNode{\Nil, $\pi_0$}\;
	\Visitor.\Traverse{$\tau_{root}$}\nllabel{alg:line:traverse}\;
	$\pi_{canon}$							\assign \CanonLeaf.$\pi$\;
	\tcp{Return just the permutation. It is then \\up to the user to permute the graph.}
	\KwRet $\overline{\pi_{canon}}$\;
}
\end{algorithm2e}
\vfill
\begin{algorithm2e}[H]\algTextSize
\caption{Tree Traversal Support}\label{alg:tt}
\FnAddTerminal{\nllabel{alg:line:addTerminal}\AddTerminal{$\tau_{leaf}$}} {
	\If{\CanonLeaf $= \Nil$} {
		\CanonLeaf $= \tau_{leaf}$\;
		\KwRet\;
	}
	$\pi_{canon}$				\assign \CanonLeaf.$\pi$\;
	$\pi_{leaf}$				\assign $\tau_{leaf}$.$\pi$\;
	$G_{canon}$					\assign $G^{\overline{\pi_{canon}}}$\;
	$G_{leaf}$					\assign $G^{\overline{\pi_{leaf}}}$\;
	\If{$G_{leaf} \repLess G_{canon}$} {
		\CanonLeaf		\assign $\tau_{canon}$\;
	}\ElseIf{$G_{leaf} \repEq G_{canon}$} {
		\Visitor.\IsomorphicLeaf{$\tau_{leaf}$}\nllabel{alg:line:isoLeaf}\;
	}
}
\BlankLine
\FnMakeChild{\nllabel{alg:line:makeChild}\MakeChild{$\tau_{parent}$, $w$}} {
	$\pi_{child}$					\assign $\tau_{parent}$.$\pi \downarrow w$\;
	$\tau_{child}$				\assign \MakeTreeNode{$\tau_{parent}$, $\pi_{child}$}\;
	\If{$\tau_{child} = \Nil$} {
		$\tau_{parent}$.\ChildPruned{$w$}		\assign \True\;
	}\Else {
		$\tau_{child}.\IndVertex$								\assign $w$\;
		$\tau_{parent}$.\Child{$w$}							\assign \NonOwningRef{$\tau_{child}$}\nllabel{alg:line:nonOwningRef}\;
	}
	\KwRet{$\tau_{child}$}\;
}
\end{algorithm2e}
\end{minipage}\hfill
\begin{minipage}[t][36em]{\algwrapRatio\textwidth}
\removelatexerror
\begin{algorithm2e}[H]\algTextSize
\caption{Tree Node}\label{alg:tree}
\FnMakeTreeNode{\MakeTreeNode{$\tau_{parent}$, $\pi$}} {
	$\tau$.\Parent						\assign $\tau_{parent}$\;
	$\tau$.$\pi$								\assign $\pi$\;
	$\tau$.\IsPruned					\assign \False\;
	\Visitor.\TreeNodeBegin{$\tau$}\nllabel{alg:line:treeBegin}\;
	\If{\Not $\tau$.\IsPruned} {
		\Visitor.\Refine{$\tau$}\nllabel{alg:line:refine}\;
		\If{\Not $\tau$.\IsPruned \And \Not $\tau$.$\pi$ discrete}{
			$\tau$.\TargetCell		\assign \Visitor.\SelectTargetCell{$\tau$}\nllabel{alg:line:targetCell}\;
			\tcp{Initialize children references \\and pruning status.}
			\ForEach{$w \in \tau$.\TargetCell} {
				$\tau$.\Child{$w$} 					\assign \Nil\;
				$\tau$.\ChildPruned{$w$}	\assign \False\;
			}
		}
	}
	\Visitor.\TreeNodeEnd{$\tau$}\nllabel{alg:line:treeEnd}\;
	\If{$\tau$.\IsPruned} {
		\KwRet \Nil\;
	}\Else {
		\KwRet $\tau$\;
	}
}
\BlankLine
\FnDestroyTreeNode{\nllabel{alg:line:destroy}\DestroyTreeNode{$\tau$}} {
	\tcp{Automatically called when \\the reference count for $\tau$ goes to 0.}
	\Visitor.\TreeNodeDestroy{$\tau$}\nllabel{alg:line:treeDestroy}\;
	\If{$\tau$.\Parent $\neq \Nil$} {
		$w$									\assign $\tau$.\IndVertex\;
		$\tau$.\Parent.\Child{$w$}			\assign \Nil\;
		$\tau$.\Parent.\ChildPruned{$w$}		\assign \True\;
	}
}
\end{algorithm2e}
\vfill
\begin{algorithm2e}[H]\algTextSize
\caption{Tree Pruning Support}\label{alg:prune}
\FnPruneTree{\nllabel{alg:line:pruneTree}\PruneTree{$\tau$}} {
	\If{$\tau$.$\pi$ is discrete} {
		\If{$\tau =$ \CanonLeaf} {
			\CanonLeaf	\assign \Nil\;
		}
		\KwRet\;
	}
	\ForEach{$w \in \tau$.\TargetCell} {
		$\tau$.\ChildPruned{$w$}		\assign \True\;
		$\tau_{child}$													\assign $\tau$.\Child{$w$}\;
		\If{$\tau_{child} \neq \Nil$} {
			\PruneTree{$\tau_{child}$}\;
		}
	}
}
\end{algorithm2e}
\end{minipage}
\caption[]{The core of the canonization algorithm framework.
Alg.~\ref{alg:core}: the entry point for canonization, called with the input graph, a compound visitor,
and objects for incorporating vertex and edge attributes.
Alg.~\ref{alg:tt}: the supporting methods for tree traversal algorithms, with \MakeChild for making new child nodes specified by the vertex to individualize,
and the method \AddTerminal for initiating leaf comparison.
Alg.~\ref{alg:tree}: the constructor and destructor methods for tree nodes.
Note that the destructor is automatically called when the reference count of a node reaches zero.
Alg.~\ref{alg:prune}: the method for marking a subtree as pruned.
Note that each core method calls specific methods on the visitors to facilitate injection of code at appropriate points.
Also, each tree node reference is an owning reference, except the one created with \NonOwningRef in \refAlgLine{tree}{nonOwningRef}.
Additionally, the C++ implementation on GitHub can be reached via the `\codeLinkText' hyperlinks in the right margin.
}
\label{fig:coreWrap}
\end{figure*}

\subsection{Weifeler-Leman Refinement and Edge Attributes}
\label{app:wl1}
The WL-1 algorithm refines a cell by distinguishing the vertices by their degree to other cells.
That is, for a cells to be refined $X\subseteq V$ and a cell to refine with $W\subseteq V$,
the degrees $d(x, W)$ for each $x\in X$ are considered.
We have generalized the WL-1 algorithm to exploit edge attributes, essentially by abstracting the degree function.
Ordinarily it is assumed to have the signature $d\colon V\times 2^V\rightarrow \mathbb{N}_0$,
but we generalize it to return a map from each possible edge attribute to the number of edges with that attribute.
That is $d$ has the signature $V\times 2^V\rightarrow (\Omega_E\rightarrow \mathbb{N}_0)$.
Such mappings are totally ordered, as we assume $\Omega_E$ is to be totally ordered.

In practice we delegate the edge handling to the given \EHandler object,
such that the user can decide the most efficient way to count edges and sort vertices.
A high-level description of the WL-1 algorithm without edge attribute handling is shown in Alg.~\ref{alg:wl}.
In this formulation the delegation to the \EHandler object happens in addition to line \ref{alg:line:wl:reset}, \ref{alg:line:wl:inc}, \ref{alg:line:wl:sort}, and \ref{alg:line:wl:split}.
\begin{algorithm2e}\algTextSize
\caption{WL-1 Refinement\hfill\codeLink{refine/WL_1.hpp}{L85}}\label{alg:wl}
\Input{$\pi$, a reference to an ordered partition}
\Input{$Q$, a non-empty FIFO queue of cells of $\pi$}
\Data{\EdgeCount, an associative array $V\rightarrow \mathbb{N}_0$, for counting neighbours}
\While{$Q$ not empty \And $\pi$ not discrete} {
	$W \assign $Q$.\FifoPop{}$\;
	\ForEach{$v\in V$} {
		$\EdgeCount{v} \assign 0$\nllabel{alg:line:wl:reset}\;
	}
	\ForEach{vertex $w\in W$} {
		\ForEach{edge $(w, x) \in E$} {
			$\EdgeCount{x}	\assign \EdgeCount{x} + 1$\nllabel{alg:line:wl:inc}\;
		}
	}
	\ForEach{cell $X\in \pi$} {
		\emph{Sort the vertices of $X$ according to the counters in $C$.}\nllabel{alg:line:wl:sort}\;
		\emph{Split $X$ into cells $X_1, X_2, \dots, X_k$ according to common values of $C$.}\nllabel{alg:line:wl:split}\;
		\emph{Report each cell split using \Visitor.\NewCell.}\;
		\If{$X\in Q$} {
			$Q$.\FifoRemove{$X$}\;
			\ForEach{$X_i \in \{X_1, X_2, \dots, X_k\}$} {
				$Q$.\FifoPush{$X_i$}\;
			}
		}\Else{
			$X_{max}\assign$ the first cell of $X_1, X_2, \dots, X_k$ of maximum cardinality\;
			\ForEach{$X_i \in \{X_1, X_2, \dots, X_k\}\backslash \{X_{max}\}$} {
				$Q$.\FifoPush{$X_i$}\;				
			}
		}
	}
}
\end{algorithm2e}

\subsection{Tree Traversal}
The tree traversal visitor uses the methods \MakeChild and \AddTerminal, Fig.~\ref{fig:coreWrap}, to create new tree nodes and report leaf nodes.
Before deciding which child to create next, it is expected to call the visitor method \BeforeDescend to allow for pruning of children.
Pruning of tree nodes is done by visitors by calling the \PruneTree procedure (\refAlgLine{tree}{pruneTree}),
which does not remove the designated subtree, but simply sets a flag \IsPruned on each node.
Visitors are then responsible for checking this flag before inspected a tree node.
An example of the DFS implementation is shown in Alg.~\ref{alg:dfs}.
\begin{figure*}
\begin{minipage}[t]{\algwrapRatio\textwidth}
\removelatexerror
\begin{algorithm2e}[H]\algTextSize
\caption{Visitor: DFS\hfill\codeLink{tree_traversal/dfs.hpp}{32}}\label{alg:dfs}
\VisFn{\Traverse{$\tau$}} {
	\Visitor.\BeforeDescend{$\tau$}\;
	\If{$\tau$.\IsPruned} {
		\KwRet\;
	}
	\If{$\tau$.$\pi$ is discrete} {
		\AddTerminal{$\tau$}\;
		\KwRet\;
	}
	\ForEach{$w\in \tau$.\TargetCell} {
		\Visitor.\BeforeDescend{$\tau$}\;
		\If{$\tau$.\IsPruned} {
			\KwRet\;
		}
		\If{$\tau$.\ChildPruned{$w$}} {\Continue\;}
		$\tau_{child}$	\assign \MakeChild{$\tau$, $w$}\;
		\If{$\tau_{child} \neq \Nil$} {
			\Traverse{$\tau_{child}$}\;
		}
	}
}
\end{algorithm2e}
\begin{algorithm2e}[H]\algTextSize
\caption{Visitor: Pruning With Automorphisms\hfill\codeLink{aut/pruner_base.hpp}{95}}\label{alg:autPruner}
\InstanceData{$A = \{\}$, a set of permutations, generating the group $\langle A\rangle$.}
\TreeNodeData{$k = 0$, number of permutations used for last pruning.}
\TreeNodeData{\Stab $= \{\}$, (a subset of) the stabilizer of $\langle A\rangle$ with respect to the tree node.}
\SetKwFunction{PruneChildren}{pruneChildren}
\Fn{\PruneChildren{$\tau$}} {
	\If{$\tau$.\Parent $= \Nil$} {
		$k_p$			\assign $|A|$\;
	}\Else{
		\PruneChildren{$\tau$.\Parent}\;
		$k_p$			\assign $\tau$.\Parent.$k$\;
	}
	\If{$\tau$.$k = k_p$} {
		\Return\;
	}	
	\emph{Update $\tau$.\Stab and $\tau$.$k$.}\;
	\emph{Prune children of $\tau$, using the permutations of $\tau$.\Stab,}\;
	\emph{while preserving \CanonLeaf.}\;
}
\BlankLine
\VisFn{\BeforeDescend{$\tau$}} {
		\PruneChildren{$\tau$}\;
}
\BlankLine
\VisFn{\IsomorphicLeaf{$\tau_a$}} {
	$\tau_c$				\assign \CanonLeaf\;
	$\pi_c$					\assign $\tau_c$.$\pi$\;
	$\pi_a$					\assign $\tau_a$.$\pi$\;
	$\gamma$			\assign	$\overline{\pi_c}\pi_a$\;
	$A$							\assign $A\cup\{\gamma\}$\;
	$\tau_p$				\assign 	\emph{The ancestor of $\tau_a$ for which its parent is \LCA{$\tau_c$, $\tau_a$}}\;
	\PruneTree{$\tau_p$}\;
}
\BlankLine
\VisFn{\ImplicitAutomorphism{$\gamma$}} {
	$A = A\cup \{\gamma\}$\;
}
\end{algorithm2e}
\end{minipage}\hfill
\begin{minipage}[t]{\algwrapRatio\textwidth}
\removelatexerror
\begin{algorithm2e}[H]\algTextSize
\caption{Visitor: Implicit Automorphisms, Cell Size 2\hfill\codeLink{aut/implicit_size_2.hpp}{18}}\label{alg:invCell2}
\InstanceData{$\gamma = (1)$, a permutation, initial the identity.}
\TreeNodeData{\Fulfilled $=\False$, whether the partition of the tree node fulfills the conditions of the lemma.}
\VisFn{\TreeNodeBegin{$\pi$}} {
	\If{$\pi$.\Parent $=$ \Nil} {\KwRet\;}
	\If{\Not $\pi$.\Parent.\Fulfilled} {\KwRet\;}
	$\pi$.\Fulfilled		\assign \True\;
	\tcp{All cells in the parent have size 1 or 2, so the cell with the individualized vertex has a singleton neighbor.
	Add a swap of those two vertices to the permutation.}
	$u$									\assign $\pi$.\IndVertex\;
	$v$									\assign \Vertex{\Pos{$u$, $\pi$}$ + 1$, $\pi$}\;
	$\gamma$					\assign $\gamma \cdot (u\ v)$\tcp*{Permutation composition.}\;
}
\BlankLine
\VisFn{\NewCell{$\pi$, $p$}} {
	\If{\Not $\pi$.\Fulfilled} {\KwRet\;}
	\tcp{All cells have size 1 or 2, so the new cell at position $p$ has a neighbor which is also a singleton.
	Add a swap of those two vertices to the permutation.}	
	$u$									\assign \Vertex{$p - 1$, $\pi$}\;
	$v$									\assign \Vertex{$p$, $\pi$}\;
	$\gamma$					\assign $\gamma \cdot (u\ v)$\;
}
\BlankLine
\VisFn{\TreeNodeEnd{$\pi$}} {
	\If{$\pi$.\IsPruned} {
		\If{$\pi$.\Fulfilled} {
			$\gamma$	\assign $(1)$\;
		}
		\KwRet\;
	}
	\tcp{Report the automorphism or check if we now fulfill the lemma.}
	\If{$\pi$.\Fulfilled} {
		\Visitor.\ImplicitAutomorphism{$\gamma$}\;
		$\gamma$		\assign $(1)$\;
	}\ElseIf{\Not $\pi$ discrete} {
		\If{all cells of $\pi$ have size 1 or 2} {
			$\pi$.\Fulfilled			\assign \True\;
		}
	}
	\tcp{Perform pruning.}
	\If{$\pi$.\Fulfilled \And \Not $\pi$ discrete} {
		\tcp{The target cell only has two vertices.}
		$u, v$		\assign $\pi$.\TargetCell\;
		\tcp{Other visitors may have already pruned children.}
		\If{\Not $\pi$.\ChildPruned{$u$} \And \Not $\pi$.\ChildPruned{$v$}} {
			$\pi$.\ChildPruned{$v$}		\assign \True\;
		}
	}
}
\end{algorithm2e}
\end{minipage}
\caption[]{
Pseudocode for a selection of visitors.
The visitor for automorphism pruning, Alg.~\ref{alg:autPruner}, is in the implementation
parameterized such that it can be used with different implementations of permutation group constructs.
Alg.~\ref{alg:invCell2} is an implementation of one of the cases in Lemma 2-25 \cite{pgi:1},
where an automorphism can be deduced when a WL-$k$ algorithm is used to refine an ordered partition containing only cells of size 1 or 2.
}
\end{figure*}

\subsection{Automorphisms}
Explicit automorphisms are provided by the core through the visitor method \IsomorphicLeaf,
where some pruning is immediately done.
Implicit automorphisms are reported from visitors through \ImplicitAutomorphism.
Pruning of children takes place in the \BeforeDescend visitor method.
Pseudocode for a generic automorphism pruner is shown in Alg.~\ref{alg:autPruner},
while pseudocode for a visitor for deducing implicit automorphisms in specific cases is shown in Alg.~\ref{alg:invCell2}.

\section{Examples of Search Tree Visualization}
\label{app:stats}
\begin{sidewaysfigure*}
\centering
\includegraphics[width=\textwidth]{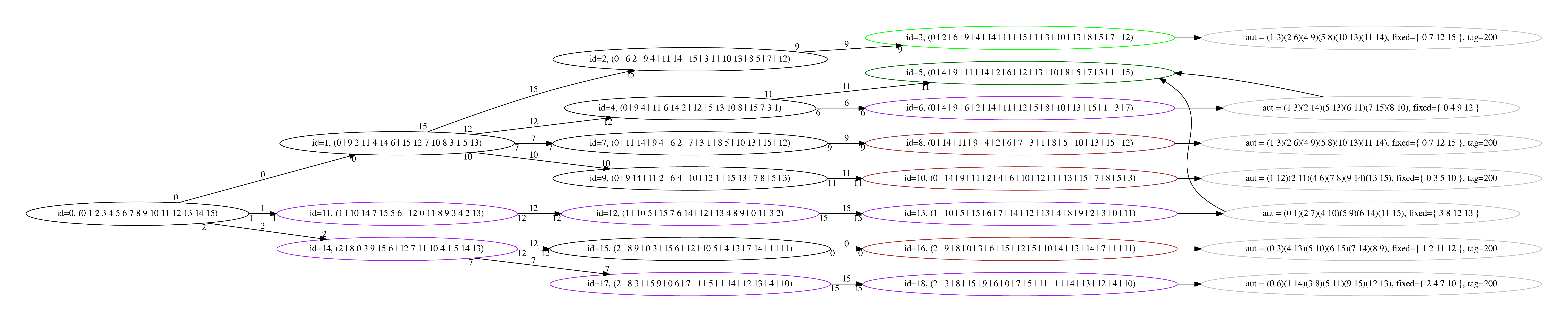}\\
\includegraphics[width=\textwidth]{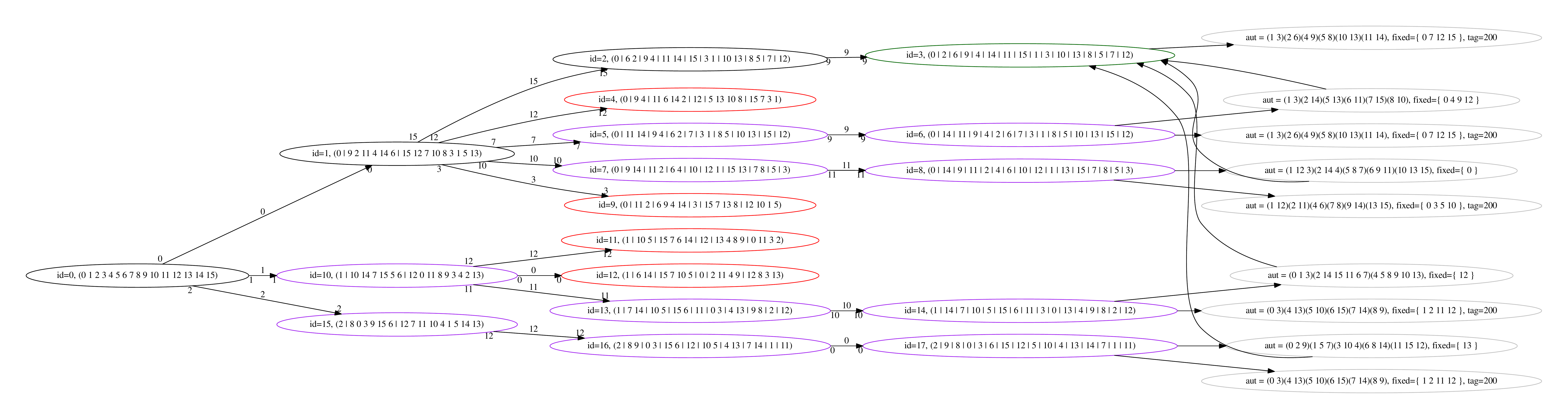}
\caption[]{
Example of search trees generated from the same graph (\graphCol{latin-4}), with the same random permutation of the input.
DFS was used for tree traversal and FLM for target cell selection. In the top tree no node invariants were enabled, while all were enabled for the bottom tree.
Each tree node has a label with an ID, indicating the order of node creation, and the associated ordered partition.
Each tree edge is labelled with the vertex being individualilzed.
The right-most gray nodes (with a label ``aut = \dots``) are not tree nodes, but represent discovered automorphisms.
Two types exist: explicit automorphisms, having an in-edge and out-edge from/to leaves indicating which permuted graphs were used to discover the automorphism,
and implicit automorphisms, having only an in-edge from the tree node where some visitor (implicitly indicated by the ``tag'') discovered it.
In this case all implicit automorphisms happened to be discovered in leaf nodes.
A red node was pruned during creation, here because of node invariants. Purple means that the node has been pruned (indirectly) through the \PruneTree method.
A brown leaf node was found worse by a comparison with the current best leaf.
A light green leaf node was once the best leaf, but was later discarded, while the dark green leaf is the resulting canonical form.
}
\label{fig:trees:small}
\end{sidewaysfigure*}

\begin{figure*}
\centering
\begin{minipage}[c][\textheight]{0.2\textwidth}
\includegraphics[height=\textheight]{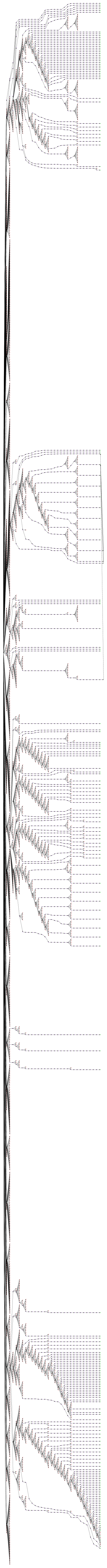}
\end{minipage}%
\begin{minipage}[c][\textheight]{0.8\textwidth}
\includegraphics[width=0.3\textwidth]{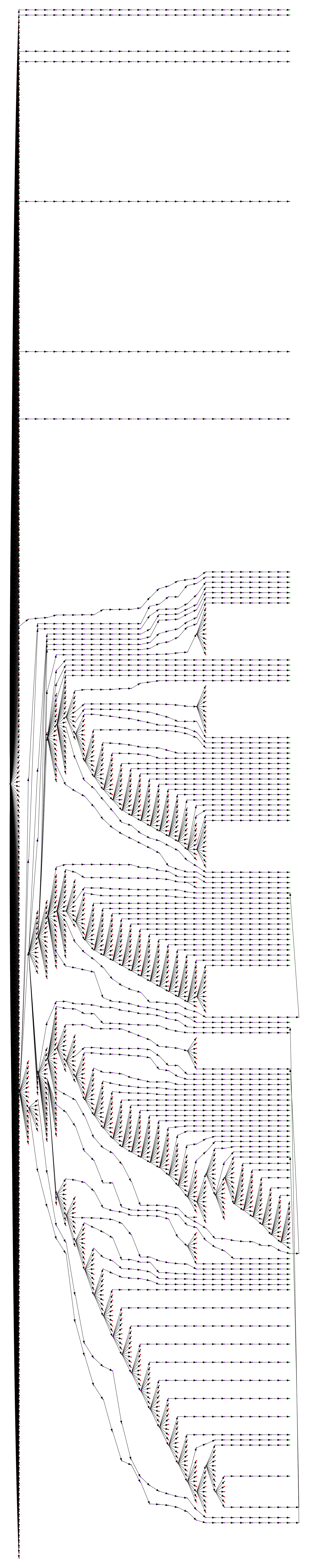}
\hfill
\includegraphics[width=0.3\textwidth]{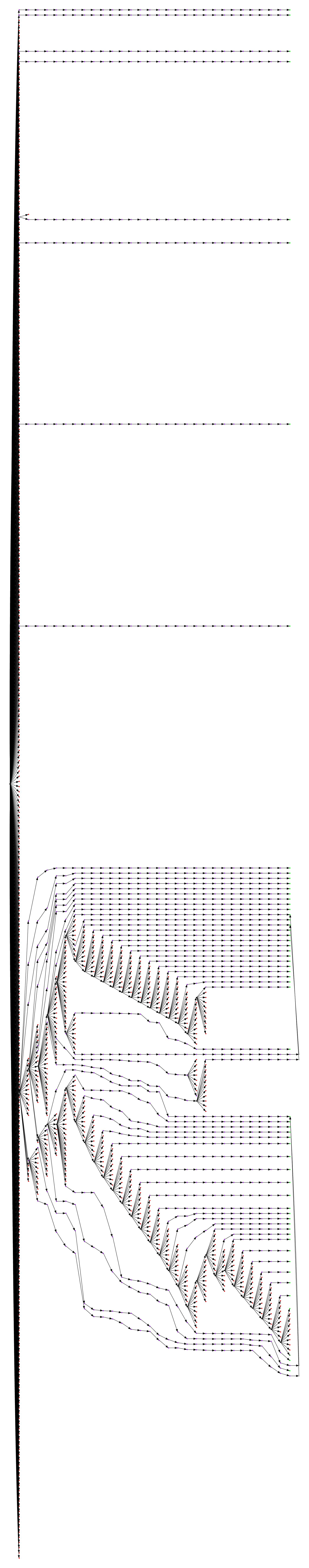}

\begin{minipage}{\textwidth}
\caption[]{
Example of search trees with different node invariants for a graph (\graphCol{f-lex-reg-10-2}) with 340 vertices.
The root of the trees are towards the left, and we have in these figures removed all labels.
Left: the search tree when only PL is enabled, 6723 tree nodes.
Middle: with PL and T enabled, 3174 nodes.
Right: with all three (PL, T, and Q) enabled, 2069 nodes.
The left tree is down-scaled compared to the middle and right trees.
The FLM target cell selector was used, and the tree traversal was BFSExp,
with the characteristic experimental paths from the tree traversal are highly visible in all three tress.
The importance of node invariants is quite noticable, as entire subtrees rooted high up in the tree are pruned when enabling more invariants.
}
\label{fig:trees:large}
\end{minipage}

\vfill
\end{minipage}
\end{figure*}

The provided stats visitor (\codeLink{visitor/stats.hpp}{14}) makes it trivial to create visualizations of the explored search tree, in addition to obtaining statistics.
An example of search tree visualization for a small graph is shown in
Fig.~\ref{fig:trees:small} and for a larger graph in
Fig.~\ref{fig:trees:large}.

\section{Investigating the Number of Allocated Nodes}
\label{sec:app:maxNumNodes}
\begin{figure*}
 \centering
\begin{tikzpicture}
\begin{axis}[
plotStyle/.style={mark size=0.2},
xlabel={Total tree nodes},
ylabel={Allocated tree nodes},
legend pos=north west, legend cell align=left,
width=\textwidth, height=7cm,
xmin=-70, xmax=2550
]
\addplot[plotStyle, green] table {figures/numNodes/dfs.txt};
\addlegendentry{DFS}
\addplot[plotStyle, teal] table {figures/numNodes/bfs-exp.txt};
\addlegendentry{BFSExp}
\addplot[plotStyle, gray] table {figures/numNodes/bfs-exp-m.txt};
\addlegendentry{BFSExpM}
\end{axis}
\end{tikzpicture}
\caption[]{
Trace of how many tree nodes were allocated every time a new tree node was created.
We clearly see BFSExpM hitting the limit of 104 tree nodes.
A large amount of nodes are later deallocated it switches back from DFS into BFSExp mode.
For this choice of memory bound we see a clear trade-off between memory and time, as BFSExpM finishes later than BFSExp.
It is however still faster than DFS.
}
\label{fig:numNodes}
\end{figure*}
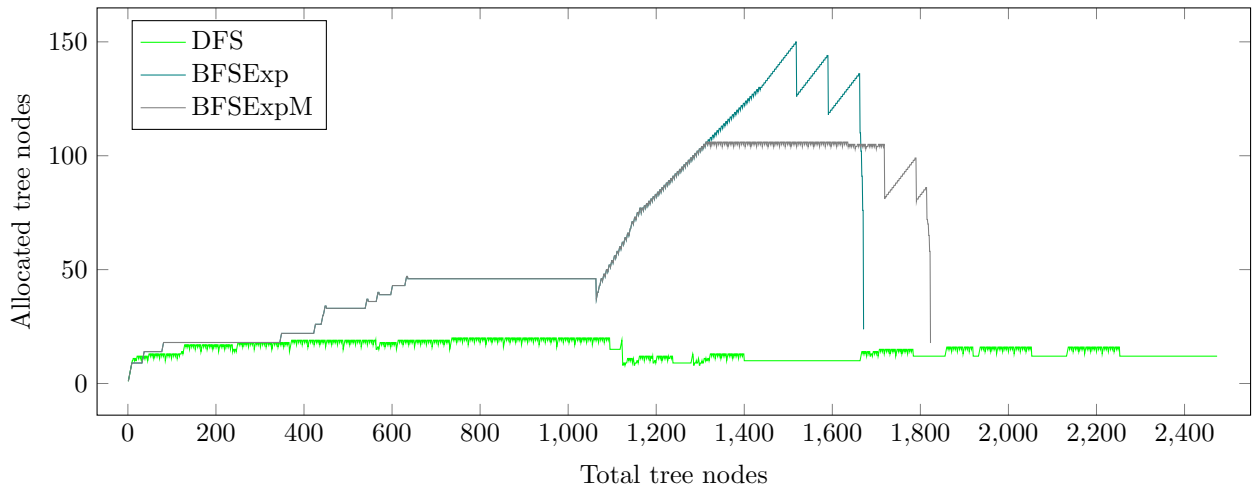

The reson for developing the memory sensitive tree traversal BFSExpM is that the plain BFSExp may have a very large amount of tree nodes allocated at the same time.
For investigating how that number develops we can use the provided debug visitor (\codeLink{visitor/debug.hpp}{15}),
which facilitates printing of log messages.
It additionally keeps track of the number of tree nodes allocated, using the \TreeNodeBegin and \TreeNodeDestroy methods.
We can thus immediately visualize how the number of allocated nodes develop.
As an example we illustrate this for a relatively small graph (\graphCol{cfi-rigid-d3-1260-04-2}, 1260 vertices).
We ran the same input permutation with DFS, BFSExp, and BFSExpM limited to \SI{2}{\mega\byte}, see Fig.~\ref{fig:numNodes}.
The current implementation of BFSExpM estimates the memory usage conservatively from the number of arrays of size $n$ in each tree node and the selected integer type.
Currently we use 32 bit integers and there are 4 arrays per tree node, thereby allowing BFSExpM 104 tree nodes before it goes into DFS mode.
Here the ordinary DFS memory usage is added, which in this case is at most 7 tree nodes extra (the maximum distance from the root for this search tree).

\end{document}